\documentclass[twocolumn]{aastex702}

\usepackage{amsmath}
\usepackage{graphicx}

\journalinfo{Draft version \today}

\begin{document}

\title{Investigating a TeV-Halo Candidate around PSR~J1928+1746 with HAWC}

\correspondingauthor{M.~Najafi}

\author[0000-0001-8749-1647]{R.~Alfaro}
\affiliation{Instituto de F\'{i}sica, Universidad Nacional Autónoma de México, Ciudad de Mexico, Mexico  }
\email{ruben@fisica.unam.mx}

\author{C.~Alvarez}
\affiliation{Universidad Autónoma de Chiapas, Tuxtla Gutiérrez, Chiapas, México}
\email{crabpulsar@hotmail.com}

\author{A. Andr\'{e}s}
\affiliation{Instituto de Astronom\'{i}a, Universidad Nacional Autónoma de México, Ciudad de Mexico, Mexico}
\email{alexis.andres1210@gmail.com}

\author{E. Anita-Rangel}
\affiliation{Instituto de Astronom\'{i}a, Universidad Nacional Autónoma de México, Ciudad de Mexico, Mexico}
\email{earangel@astro.unam.mx}

\author[0000-0002-0595-9267]{M.~Araya}
\affiliation{Universidad de Costa Rica, San José 2060, Costa Rica}
\email{miguel.araya@ucr.ac.cr}

\author{J.C.~Arteaga-Velázquez}
\affiliation{Universidad Michoacana de San Nicolás de Hidalgo, Morelia, Mexico }
\email{juan.arteaga@umich.mx}

\author[0000-0002-4020-4142]{D.~Avila Rojas}
\affiliation{Instituto de Astronom\'{i}a, Universidad Nacional Autónoma de México, Ciudad de Mexico, Mexico}
\email{doavila@astro.unam.mx}

\author[0000-0002-2084-5049]{H.A.~Ayala Solares}
\affiliation{Temple University, Department of Physics, 1925 N. 12th Street, Philadelphia, PA 19122, USA}
\email{hugo.ayala.solares@temple.edu}

\author[0000-0002-5529-6780]{R.~Babu}
\affiliation{Department of Physics and Astronomy, Michigan State University, East Lansing, MI, USA}
\email{baburish@msu.edu}

\author[0000-0002-3886-3739]{P.~Bangale}
\affiliation{Temple University, Department of Physics, 1925 N. 12th Street, Philadelphia, PA 19122, USA}
\email{priyadarshini.bangale@temple.edu}

\author{A.~Bernal}
\affiliation{Instituto de Astronom\'{i}a, Universidad Nacional Autónoma de México, Ciudad de Mexico, Mexico}
\email{abel@astro.unam.mx}

\author[0000-0002-4042-3855]{K.S.~Caballero-Mora}
\affiliation{Universidad Autónoma de Chiapas, Tuxtla Gutiérrez, Chiapas, México}
\email{karen.scm@gmail.com}

\author[0000-0003-2158-2292]{T.~Capistran}
\affiliation{Instituto Nacional de Astrof\'{\i}sica, \'{O}ptica y Electr\'{o}nica, Tonantzintla, Puebla, MX 72840}
\email{tcapistranc@gmail.com}

\author[0000-0002-8553-3302]{A.~Carrami\~{n}ana}
\affiliation{Instituto Nacional de Astrof\'{\i}sica, \'{O}ptica y Electr\'{o}nica, Tonantzintla, Puebla, MX 72840}
\email{alberto@inaoep.mx}

\author{F.~Carreón}
\affiliation{Instituto de Astronom\'{i}a, Universidad Nacional Autónoma de México, Ciudad de Mexico, Mexico}
\email{mfcarreong@gmail.com}

\author[0000-0002-6144-9122]{S.~Casanova}
\affiliation{Instytut Fizyki Jadrowej im Henryka Niewodniczanskiego Polskiej Akademii Nauk, IFJ-PAN, Krakow, Poland }
\email{sabrinacasanova@gmail.com}

\author[0000-0002-7607-9582]{U.~Cotti}
\affiliation{Universidad Michoacana de San Nicolás de Hidalgo, Morelia, Mexico }
\email{umberto.cotti@umich.mx}

\author[0000-0002-1132-871X]{J.~Cotzomi}
\affiliation{Facultad de Ciencias F\'{i}sico Matemáticas, Benem\'{e}rita Universidad Aut\'{o}noma de Puebla, Puebla, M\'{e}xico }
\email{jcotzomi@yahoo.com.mx}

\author[0000-0001-9643-4134]{E.~De la Fuente}
\affiliation{Departamento de F\'{i}sica, Centro Universitario de Ciencias Exactas e Ingenierias, Universidad de Guadalajara, Guadalajara, M\'{e}xico }
\email{edfuente@gmail.com}

\author{P.~Desiati}
\affiliation{Department of Physics, University of Wisconsin-Madison, Madison, WI 53706, USA }
\email{paolo.desiati@icecube.wisc.edu}

\author[0000-0002-7574-1298]{N.~Di Lalla}
\affiliation{Department of Physics, Stanford University: Stanford, CA 94305–4060, USA}
\email{niccolo.dilalla@stanford.edu}

\author{R.~Diaz Hernandez}
\affiliation{Instituto Nacional de Astrof\'{\i}sica, \'{O}ptica y Electr\'{o}nica, Tonantzintla, Puebla, MX 72840}
\email{dihera77@gmail.com}

\author[0000-0002-2987-9691]{M.A.~DuVernois}
\affiliation{Department of Physics, University of Wisconsin-Madison, Madison, WI 53706, USA }
\email{duvernois@icecube.wisc.edu}

\author[0000-0002-0087-0693]{J.C.~Díaz-Vélez}
\affiliation{Department of Physics, University of Wisconsin-Madison, Madison, WI 53706, USA }
\email{juancarlos@icecube.wisc.edu}

\author[0000-0001-5737-1820]{K.~Engel}
\affiliation{Department of Physics, University of Maryland, College Park, MD, USA }
\email{kristi.engel23@gmail.com}

\author[0000-0003-2423-4656]{T.~Ergin}
\affiliation{Department of Physics and Astronomy, Michigan State University, East Lansing, MI, USA}
\email{ergin.tulun@gmail.com}

\author[0000-0001-7074-1726]{C.~Espinoza}
\affiliation{Instituto de F\'{i}sica, Universidad Nacional Autónoma de México, Ciudad de Mexico, Mexico }
\email{m.catalina@fisica.unam.mx}

\author[0000-0002-0173-6453]{N.~Fraija}
\affiliation{Instituto de Astronom\'{i}a, Universidad Nacional Autónoma de México, Ciudad de Mexico, Mexico}
\email{nifraija@astro.unam.mx}

\author[]{S.~Fraija}
\affiliation{Instituto de Astronom\'{i}a, Universidad Nacional Autónoma de México, Ciudad de Mexico, Mexico}
\email{sarafraija@hotmail.com}

\author[0000-0002-4188-5584]{J.A.~García-González}
\affiliation{Tecnologico de Monterrey, Escuela de Ingenier\'{i}a y Ciencias, Ave. Eugenio Garza Sada 2501, Monterrey, N.L., Mexico, 64849}
\email{anteus79@tec.mx}

\author[0000-0003-1122-4168]{F.~Garfias}
\affiliation{Instituto de Astronom\'{i}a, Universidad Nacional Autónoma de México, Ciudad de Mexico, Mexico}
\email{fergar@astro.unam.mx}

\author{N.~Ghosh}
\affiliation{Department of Physics, Michigan Technological University, Houghton, MI, USA }
\email{nghosh1@mtu.edu}

\author{S.~Giri}
\affiliation{Instytut Fizyki Jadrowej im Henryka Niewodniczanskiego Polskiej Akademii Nauk, IFJ-PAN, Krakow, Poland }
\email{suvendu.giri@ifj.edu.pl}

\author[0000-0002-5209-5641]{M.M.~Gonz\'{a}lez}
\affiliation{Instituto de Astronom\'{i}a, Universidad Nacional Autónoma de México, Ciudad de Mexico, Mexico}
\email{magda@astro.unam.mx}

\author{J.A.~González}
\affiliation{Universidad Michoacana de San Nicolás de Hidalgo, Morelia, Mexico }
\email{jose.gonzalez.c@umich.mx}

\author[0000-0002-9790-1299]{J.A.~Goodman}
\affiliation{Department of Physics, University of Maryland, College Park, MD, USA }
\email{goodman@umd.edu}

\author{S.~Groetsch}
\affiliation{Department of Physics, Michigan Technological University, Houghton, MI, USA }
\email{sjgroets@mtu.edu}

\author{J.~Gyeong}
\affiliation{Department of Physics, Sungkyunkwan University, Suwon 16419, South Korea}
\email{kyoungjh1011@naver.com}

\author[0000-0001-9844-2648]{J.P.~Harding}
\affiliation{Los Alamos National Laboratory, Los Alamos, NM, USA }
\email{jpharding@lanl.gov}

\author[0000-0001-5169-723X]{I.~Herzog}
\affiliation{Department of Physics and Astronomy, Michigan State University, East Lansing, MI, USA}
\email{herzogia@msu.edu}

\author{D.~Huang}
\affiliation{Department of Physics and Astronomy and the Bartol Research Institute, University of Delaware, Newark, DE 19716, USA}
\email{dezhih@mtu.edu}

\author[0000-0002-5527-7141]{F.~Hueyotl-Zahuantitla}
\affiliation{Universidad Autónoma de Chiapas, Tuxtla Gutiérrez, Chiapas, México}
\email{filihz@gmail.com}

\author[0000-0002-3302-7897]{P.~Hüntemeyer}
\affiliation{Department of Physics, Michigan Technological University, Houghton, MI, USA }
\email{petra@mtu.edu}

\author[0000-0001-5811-5167]{A.~Iriarte}
\affiliation{Instituto de Astronom\'{i}a, Universidad Nacional Autónoma de México, Ciudad de Mexico, Mexico}
\email{airiarte@astro.unam.mx}

\author{S.~Kaufmann}
\affiliation{Universidad Politecnica de Pachuca, Pachuca, Hgo, M\'{e}xico }
\email{skaufmann13@googlemail.com}

\author[0000-0003-4785-0101]{D.~Kieda}
\affiliation{Department of Physics and Astronomy, University of Utah, Salt Lake City, UT, USA }
\email{dave.kieda@utah.edu}

\author{R.~Lake}
\affiliation{Max-Planck Institute for Nuclear Physics, 69117 Heidelberg, Germany}
\email{rhiannon.lake@mpi-hd.mpg.de}

\author{K.~Leavitt}
\affiliation{Department of Physics, Michigan Technological University, Houghton, MI, USA }
\email{kleavitt@mtu.edu}

\author[0000-0001-5516-4975]{H.~Le\'{o}n Vargas}
\affiliation{Instituto de F\'{i}sica, Universidad Nacional Autónoma de México, Ciudad de Mexico, Mexico }
\email{hleonvar@fisica.unam.mx}

\author[0000-0001-8825-3624]{A.L. Longinotti}
\affiliation{Instituto de Astronom\'{i}a, Universidad Nacional Autónoma de México, Ciudad de Mexico, Mexico}
\email{alonginotti@astro.unam.mx}

\author[0000-0003-2810-4867]{G.~Luis-Raya}
\affiliation{Universidad Politecnica de Pachuca, Pachuca, Hgo, M\'{e}xico }
\email{gilura6969@hotmail.com}

\author[0000-0001-8088-400X]{K.~Malone}
\affiliation{Los Alamos National Laboratory, Los Alamos, NM, USA }
\email{kmalone@lanl.gov}

\author[0000-0001-9052-856X]{O.~Martinez}
\affiliation{Facultad de Ciencias F\'{i}sico Matemáticas, Benem\'{e}rita Universidad Aut\'{o}noma de Puebla, Puebla, M\'{e}xico }
\email{omartin@fcfm.buap.mx}

\author[0000-0002-2824-3544]{J.~Mart\'{i}nez-Castro}
\affiliation{Centro de Investigaci\'on en Computaci\'on, Instituto Polit\'ecnico Nacional, Mexico City, Mexico.}
\email{macj@cic.ipn.mx}

\author[0000-0001-7714-0704]{H.~Martínez-Huerta}
\affiliation{Departamento de F\'isica y Matem\'aticas, Universidad de Monterrey, Av. Morones Prieto 4500, San Pedro Garza Garc\'ia 66238, Nuevo Le\'on, M\'exico}
\email{humberto.martinezhuerta@udem.edu}

\author[0000-0002-2610-863X]{J.A.~Matthews}
\affiliation{Dept of Physics and Astronomy, University of New Mexico, Albuquerque, NM, USA }
\email{johnm@unm.edu}

\author[0000-0002-8390-9011]{P.~Miranda-Romagnoli}
\affiliation{Universidad Aut\'{o}noma del Estado de Hidalgo, Pachuca, M\'{e}xico }
\email{pa.miranda.r@gmail.com}

\author{P.E. Mir\'{o}n-Enriquez}
\affiliation{Instituto de Astronom\'{i}a, Universidad Nacional Autónoma de México, Ciudad de Mexico, Mexico}
\email{pelimi92@gmail.com}

\author[0000-0001-9361-0147]{J.A.~Morales-Soto}
\affiliation{Universidad Michoacana de San Nicolás de Hidalgo, Morelia, Mexico}
\email{jmoralessg@gmail.com}

\author[0000-0002-1114-2640]{E.~Moreno}
\affiliation{Facultad de Ciencias F\'{i}sico Matemáticas, Benem\'{e}rita Universidad Aut\'{o}noma de Puebla, Puebla, M\'{e}xico }
\email{emoreno@fcfm.buap.mx}

\author[0000-0002-7675-4656]{M.~Mostaf\'{a}}
\affiliation{School of Science \& Engineering,
University of Missouri--Kansas City,
Kansas City, MO, USA}
\email{miguel.mostafa@umkc.edu}

\author[0000-0002-7576-4799]{M.~Najafi}
\affiliation{Department of Physics, Michigan Technological University, Houghton, MI, USA }
\email[show]{mnajafi@mtu.edu}

\author[0000-0003-1059-8731]{L.~Nellen}
\affiliation{Instituto de Ciencias Nucleares, Universidad Nacional Aut\'{o}noma de M\'{e}xico, Ciudad de M\'{e}xico, M\'{e}xico }
\email{lukas@nucleares.unam.mx}

\author[0000-0002-6859-3944]{M.U.~Nisa}
\affiliation{Department of Physics and Astronomy, Michigan State University, East Lansing, MI, USA}
\email{nisamehr@msu.edu}

\author[0000-0001-7099-108X]{R.~Noriega-Papaqui}
\affiliation{Universidad Aut\'{o}noma del Estado de Hidalgo, Pachuca, M\'{e}xico }
\email{ropapaqui@gmail.com}

\author[0009-0009-2481-6921]{M. Osorio-Archila}
\affiliation{Instituto de Astronom\'{i}a, Universidad Nacional Autónoma de México, Ciudad de Mexico, Mexico}
\email{jmosorio@astro.unam.mx}

\author{N. Omodei}
\affiliation{Department of Physics, Stanford University: Stanford, CA 94305–4060, USA}
\email{nicola.omodei@stanford.edu}

\author{E.~Ponce}
\affiliation{Facultad de Ciencias F\'{i}sico Matemáticas, Benem\'{e}rita Universidad Aut\'{o}noma de Puebla, Puebla, M\'{e}xico }
\email{eponce@fcfm.buap.mx}

\author[0000-0002-8774-8147]{Y.~Pérez Araujo}
\affiliation{Instituto de F\'{i}sica, Universidad Nacional Autónoma de México, Ciudad de Mexico, Mexico }
\email{yuniorpy@gmail.com}

\author[0000-0001-5998-4938]{E.G.~Pérez-Pérez}
\affiliation{Universidad Politecnica de Pachuca, Pachuca, Hgo, M\'{e}xico }
\email{egperezp@yahoo.com.mx}

\author{S. Recchia}
\affiliation{Instytut Fizyki Jadrowej im Henryka Niewodniczanskiego Polskiej Akademii Nauk, IFJ-PAN, Krakow, Poland }
\email{sarah.recchia@ifj.edu.pl}

\author[0000-0002-6524-9769]{C.D.~Rho}
\affiliation{Department of Physics, Sungkyunkwan University, Suwon 16419, South Korea}
\email{no397@naver.com}

\author[0000-0003-1327-0838]{D.~Rosa-Gonz\'{a}lez}
\affiliation{Instituto Nacional de Astrof\'{\i}sica, \'{O}ptica y Electr\'{o}nica, Tonantzintla, Puebla, MX 72840}
\email{danrosa@inaoep.mx}

\author[0000-0002-4204-5026]{M.~Roth}
\affiliation{Los Alamos National Laboratory, Los Alamos, NM, USA }
\email{mattroth@lanl.gov}

\author[0000-0001-6079-2722]{A.~Sandoval}
\affiliation{Instituto de F\'{i}sica, Universidad Nacional Autónoma de México, Ciudad de Mexico, Mexico  }
\email{asandoval@fisica.unam.mx}

\author[0000-0001-8644-4734]{M.~Schneider}
\affiliation{Department of Physics, University of Maryland, College Park, MD, USA }
\email{mschnei4@umd.edu}

\author{J.~Serna-Franco}
\affiliation{Instituto de F\'{i}sica, Universidad Nacional Autónoma de México, Ciudad de Mexico, Mexico  }
\email{j_serna@ciencias.unam.mx}

\author{M.~Shin}
\affiliation{Department of Physics, Sungkyunkwan University, Suwon 16419, South Korea}
\email{minjishin23@gmail.com}

\author[0000-0002-1012-0431]{A.J.~Smith}
\affiliation{Department of Physics, University of Maryland, College Park, MD, USA }
\email{asmith8@umd.edu}

\author{Y.~Son}
\affiliation{Department of Physics, Sungkyunkwan University, Suwon 16419, South Korea}
\email{youngwan.son@cern.ch}

\author[0000-0002-1492-0380]{R.W.~Springer}
\affiliation{Department of Physics and Astronomy, University of Utah, Salt Lake City, UT, USA }
\email{wayne.springer@utah.edu}

\author{O.~Tibolla}
\affiliation{Universidad Politecnica de Pachuca, Pachuca, Hgo, M\'{e}xico }
\email{omar.tibolla@gmail.com}

\author[0000-0001-9725-1479]{K.~Tollefson}
\affiliation{Department of Physics and Astronomy, Michigan State University, East Lansing, MI, USA}
\email{tollefson@pa.msu.edu}

\author[0000-0002-1689-3945]{I.~Torres}
\affiliation{Instituto Nacional de Astrof\'{\i}sica, \'{O}ptica y Electr\'{o}nica, Tonantzintla, Puebla, MX 72840}
\email{ibrahim.torres23@gmail.com}

\author[0000-0002-7102-3352]{R.~Torres-Escobedo}
\affiliation{Tsung-Dao Lee Institute \& School of Physics and Astronomy, Shanghai Jiao Tong University, 800 Dongchuan Rd, Shanghai, SH 200240, China}
\email{torresramiro350@gmail.com}

\author[0000-0003-0715-7513]{E.~Varela}
\affiliation{Facultad de Ciencias F\'{i}sico Matemáticas, Benem\'{e}rita Universidad Aut\'{o}noma de Puebla, Puebla, M\'{e}xico  }
\email{enrique.varela@correo.buap.mx}

\author[0000-0001-6876-2800]{L.~Villaseñor}
\affiliation{Facultad de Ciencias F\'{i}sico Matemáticas, Benem\'{e}rita Universidad Aut\'{o}noma de Puebla, Puebla, M\'{e}xico }
\email{lvillasen@gmail.com}

\author[0000-0001-6798-353X]{X.~Wang}
\affiliation{Department of Physics, Missouri University of Science and Technology, Rolla, MO, US}
\email{xwang32@mtu.edu}

\author{Z.~Wang}
\affiliation{Department of Physics, Missouri University of Science and Technology, Rolla, MO, US}
\email{zhen@umd.edu}

\author[0000-0003-2141-3413]{I.J.~Watson}
\affiliation{University of Seoul, Seoul, Rep. of Korea}
\email{ian.james.watson@cern.ch}

\author[0009-0005-7243-1402]{H.~Wu}
\affiliation{Department of Physics, University of Wisconsin-Madison, Madison, WI 53706, USA }
\email{hwu298@wisc.edu}

\author[0009-0006-3520-3993]{S.~Yu}
\affiliation{Department of Physics, Pennsylvania State University, University Park, PA, USA}
\email{sjy5345@psu.edu}

\author[0000-0002-9307-0133]{S.~Yun-Cárcamo}
\affiliation{Drexel University, Philadelphia, PA, USA}
\email{sy645@drexel.edu}

\author{X.~Zhang}
\affiliation{Instytut Fizyki Jadrowej im Henryka Niewodniczanskiego Polskiej Akademii Nauk, IFJ-PAN, Krakow, Poland }
\email{xiyingzhangxyz@gmail.com}

\author[0000-0003-0513-3841]{H.~Zhou}
\affiliation{Tsung-Dao Lee Institute \& School of Physics and Astronomy, Shanghai Jiao Tong University, 800 Dongchuan Rd, Shanghai, SH 200240, China}
\email{hao_zhou@sjtu.edu.cn}

\author[0000-0002-8528-9573]{C.~de Le\'{o}n}
\affiliation{Universidad Michoacana de San Nicolás de Hidalgo, Morelia, Mexico  }
\email{cederik.de.leon@umich.mx}

\collaboration{all}{The HAWC Collaboration}

\begin{abstract}
We present an updated High-Altitude Water Cherenkov (HAWC) analysis of the crowded Galactic-plane area near PSR~J1928+1746, which was previously linked to the extended source 3HWC~J1928+178. With 2904 days of HAWC data, improved event reconstruction, and a multicomponent likelihood model that includes six TeV source components and Galactic
diffuse emission, we describe the region using six components and find a distinct, moderately extended source close to PSR~J1928+1746. This source, HAWC~J1928+1750, is detected from about 2~TeV to above 100~TeV and is well-described by a log-parabola spectrum with normalization $K=5.0\times10^{-15}\,\mathrm{TeV^{-1}\,cm^{-2}\,s^{-1}}$ at 12~TeV, index $\alpha=2.09$, curvature $\beta=0.35$, and Gaussian extension $\sigma=0.19^\circ$. Interpreting the emission using diffusion-halo templates under the assumption $B=8~\mu{\rm G}$ yields the diffusion coefficient $D(100~{\rm TeV})=2.1\times10^{27}\,\mathrm{cm^2\,s^{-1}}$, which is about 500 times smaller than typical interstellar diffusion values at similar energies. This suggests that particle transport around PSR~J1928+1746 is strongly suppressed under the TeV-halo interpretation. The TeV spectrum is consistent with leptonic inverse-Compton emission originating from pulsar-injected electrons and positrons, which requires less than approximately $1\%$ of the pulsar's characteristic spin-down energy. However, a hadronic contribution from the surrounding environment cannot be excluded. 

\end{abstract}

\section{Introduction}

\begin{figure*}[t!]
\centering
\gridline{
\fig{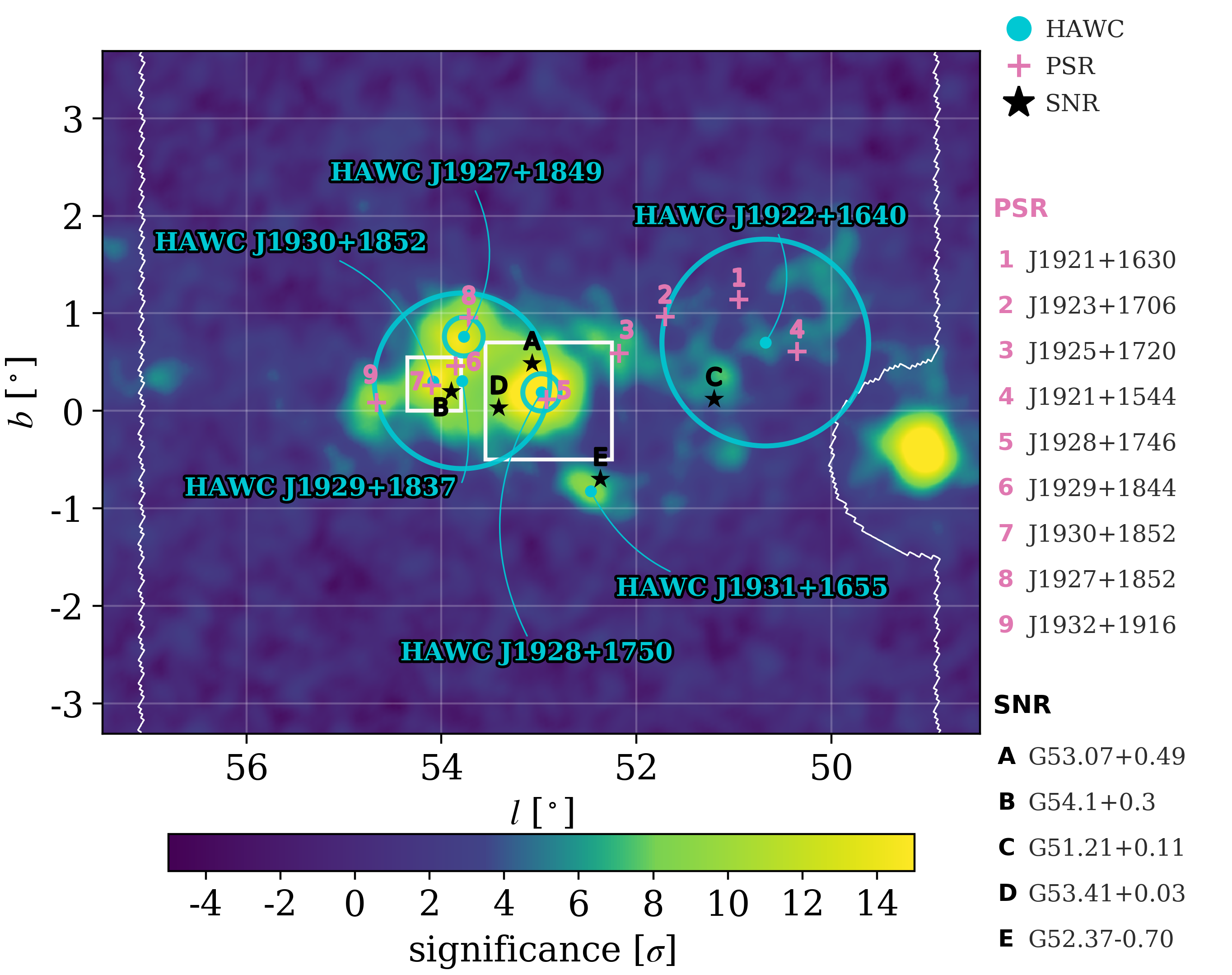}
    {0.52\textwidth}{(a)}
\fig{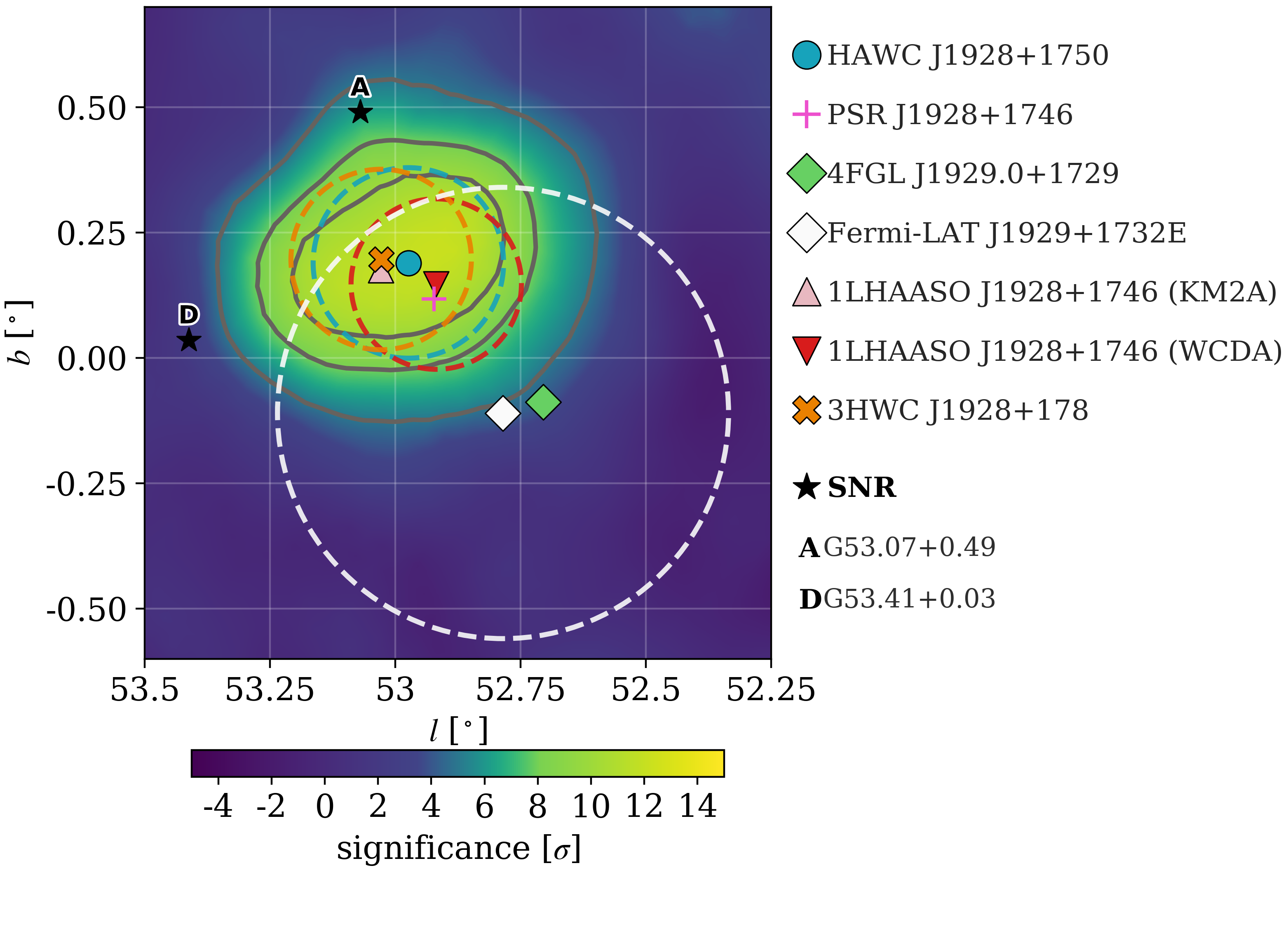}
    {0.48\textwidth}{(b)}
}
\caption{ HAWC significance maps of the J1928 region in Galactic coordinates. Panel~(a) shows the wider analysis area with the final global-model components overlaid. Blue points mark the best-fit centroids of the HAWC components, while blue circles have radii equal to the Gaussian extension parameter for extended components. Point-like components are shown without extension circles; the fitted extensions and their statistical uncertainties are listed in Table~\ref{tab:global_model}. Magenta crosses mark selected ATNF pulsars near the fitted components or relevant to their interpretation \citep{Manchester2005}, with numbered labels corresponding to the list at right; and black stars mark known or candidate SNRs, including the THOR candidate G51.21+0.11 \citep{Green2019SNR, Anderson2017THOR}. The white outline marks the visible portion of the likelihood region of interest (RoI); the full RoI spans $49^\circ<\ell<57^\circ$ and $|b|<6^\circ$, beyond the latitude range shown. The bright W51 complex is not included in the fit. The smaller white box around HAWC J1930+1852 marks the region shown in the inset of Figure~\ref{fig:j1930_sed}, while the larger white box around HAWC J1928+1750 marks the region enlarged in panel (b). Panel~(b) shows a source-only residual significance map centered on PSR~J1928+1746, obtained after subtracting the other fitted TeV components and Galactic diffuse emission; the remaining emission is dominated by HAWC~J1928+1750. The solid gray contours trace significance levels of $5\sigma$, $8\sigma$, and $10\sigma$, from outermost to innermost. Overlaid markers indicate the positions of 3HWC~J1928+178 \citep{Albert2020_3HWC, Albert2023J1928}, 1LHAASO~J1928+1746u \citep{Cao2024LHAASO}, nearby SNR candidates \citep{Anderson2017THOR}, the extended LAT component J1929+1732E from \citet{Liu2026FermiJ1928}, and the official 4FGL source 4FGL~J1929.0+1729 \citep{Ballet2023_4FGLDR4}. J1929+1732E is a designation specific to the Liu et al. analysis, not an official 4FGL name. The dashed white circle marks the \textit{Fermi}-LAT $68\%$ containment radius of J1929+1732E, $r_{68}=0.45^\circ$.}
\label{fig:roi_context}
\end{figure*}

TeV halos are spatially extended regions of very-high-energy gamma-ray emission observed around middle-aged pulsars. Halos are commonly believed to arise from inverse-Compton (IC) scattering by relativistic electrons and positrons that have left the pulsar wind nebula (PWN) and entered the nearby interstellar medium (ISM). Since firm detections of such extended TeV emission around Geminga and Monogem pulsars by HAWC, TeV halos have become a key way to study how particles propagate near pulsars and the possible contribution of pulsars to the Galactic high-energy cosmic-ray lepton population \citep{Abeysekara2017}.

These systems are especially interesting because TeV halos are characterized by particle diffusion that is strongly suppressed relative to the average ISM \citep{Abeysekara2017}. This affects how we understand the particle injection of the pulsar and whether pulsars could explain much of the observed cosmic-ray positron excess \citep{Albert2024GemMon}.
Recent HAWC results also suggest that TeV halos might be common around middle-aged pulsars  \citep{Albert2025Common}. However, it is still unclear how much their properties depend on the local environment or the source's history. 

PSR~J1928+1746 is a middle-aged, energetic pulsar with a characteristic age
$\tau_c \simeq 82~{\rm kyr}$ and a spin-down power
$\dot{E}=1.6 \times 10^{36}$~erg~s$^{-1}$ \citep{cordes2006}. 
An earlier HAWC analysis modeled the emission around this pulsar with the extended component 3HWC~J1928+178 and a broader component, HAWC~J1928-EXT \citep{Albert2023J1928}. The region also includes 3HWC~J1930+188, which is linked to the PSR~J1930+1852 and SNR~G54.1+0.3 complex, as well as the nearby source
HAWC~J1932+192. The origin of the emission near PSR~J1928+1746 remained uncertain because no clear X-ray counterpart has been found for the pulsar-centered TeV emission, and molecular material near the source could allow for hadronic interactions \citep{Albert2023J1928}.
In this paper, we refer to this Galactic-plane field, which includes PSR~J1928+1746, the G54.1+0.3 complex and the nearby TeV emission, as the J1928 region. Several ATNF pulsars are located near the fitted TeV components in this field, which introduces potential source confusion and highlights the importance of considering possible pulsar associations in regional modeling.
Figure~\ref{fig:roi_context}(a) presents the broader J1928 region, while Figure~\ref{fig:roi_context}(b) highlights the central area around PSR~J1928+1746, including the nearby 3HWC, LHAASO, and \textit{Fermi}-LAT positions discussed in this study.

A recent dedicated \textit{Fermi}-LAT analysis of this region identified an extended GeV component called J1929+1732E, using a source-search analysis that goes beyond the standard 4FGL-DR4 catalog model \citep{Liu2026FermiJ1928}. This component overlaps with the official 4FGL catalog source, 4FGL~J1929.0+1729 \citep{Ballet2023_4FGLDR4}, and lies in the same GeV to TeV complex as the previously reported HAWC emission near 3HWC~J1928+178 \citep{Albert2023J1928} and the LHAASO sources, 1LHAASO~J1928+1746u \citep{Cao2024LHAASO}. Because the \textit{Fermi}-LAT component is broader than HAWC J1928+178, with a reported $68\%$ containment radius of $r_{68}=0.45^\circ$, it likely traces a broader GeV environment, rather than only a pulsar-centered TeV component.
This region has overlapping TeV sources, possible pulsar associations, molecular material, and structured emission from the Galactic plane. Because HAWC has an energy-dependent angular resolution, with $68\%$ containment angles of order
0.3° at a few TeV and less than 0.2° at energies above tens of TeV, emission
from nearby sources can overlap in the likelihood fit
\citep{Albert2024HAWCPerformance}.
To understand the TeV excess centered on PSR J1928+1746, we need a model that accounts for nearby sources and diffuse emission in a consistent way. This helps ensure that flux from the surrounding area is not incorrectly assigned to PSR~J1928+1746. In this study, we take another look at the region using updated event reconstruction, improved gamma--hadron separation, better energy estimation, and a clear template for Galactic diffuse emission.

We revisit the J1928 region using 2904 days of HAWC data, as described in Section~\ref{sec:data}, along with the event reconstruction. Using the analysis procedure in Section~\ref{sec:method}, we first construct a global model of the region that includes nearby TeV sources and Galactic diffuse emission. The resulting regional decomposition and comparison with previous TeV measurements are presented in Section~\ref{sec:global_model}. We then focus on the extended emission near PSR~J1928+1746 in Section~\ref{sec:halo}. There, we study its morphology, spectrum, radial surface-brightness profile, and diffusion-halo interpretation. Section~\ref{sec:scenarios} discusses possible leptonic and hadronic gamma-ray production scenarios, and Section~\ref{conclusion} summarizes the main findings.

\section{HAWC Data Set and Event Selection}
\label{sec:data}

The High-Altitude Water Cherenkov (HAWC) observatory is a wide-field air-shower detector located in Mexico at an altitude of
4100~m.
Its main array has 300 water Cherenkov detectors, each with four photomultiplier tubes (PMTs). A detailed description of the HAWC instrument is provided by \citet{Abeysekara2023HAWCdetector}. HAWC detects extensive air showers initiated by gamma rays and cosmic rays with energies from several hundred GeV to several hundred TeV. 

We use 2904 days of HAWC data between June 2015 and January 2024. The likelihood analysis is performed in a region of interest (RoI) that covers
$49^\circ < \ell < 57^\circ$ and $|b| < 6^\circ$. This RoI includes emission around PSR~J1928+1746, nearby TeV sources, and large-scale Galactic diffuse emission (GDE) that could affect how sources are separated. Figure~\ref{fig:roi_context}(a) shows the analysis region, with the white boundary marking the RoI used in the likelihood fit. The bright W51 complex is excluded from the fit to avoid contamination from its structured emission.

A key step in event selection is separating air showers caused by gamma rays from the much larger background of hadronic cosmic rays. We use the standard HAWC gamma--hadron separation method \citep{Alfaro2022GH}, but replace standard cuts with cuts generated by the HAWC machine-learning (ML) classifier. These ML cuts improve gamma-ray sensitivity in the TeV band; for the Crab Nebula, they increase the detection significance by about $19\%$ compared with the standard cuts \citep{Alfaro2025MLGH}.

Previous HAWC studies of this region used $f_{\rm Hit}$, the fraction of active PMTs that recorded a signal, as a proxy for shower size and energy \citep{Abeysekara2017Crab}. In this analysis, gamma-ray energies are reconstructed using either the ground-parameter (GP) estimator, which relies on the shower charge density at ground level, or the neural-network (NN) estimator, which combines several reconstructed shower properties. The NN estimator offers better energy resolution across most of the energy range and is less biased at low energies, where threshold effects are strongest \citep{Abeysekara2019Crab}. We use the NN energy estimator in the reference analysis and the GP estimator as a systematic cross-check.

\section{Analysis Method}
\label{sec:method}

We study the J1928 region with the Multi-Mission Maximum Likelihood framework
(\texttt{3ML}) together with the HAWC Accelerated Likelihood (HAL) plugin
\citep{Vianello2015, Abeysekara2022HAL}. The HAWC data are modeled with a
binned Poisson likelihood, in which the predicted counts in each spatial and energy bin are calculated as the sum of source components and Galactic diffuse emission. We first define a reference diffuse emission component for the region. Then, we use an iterative, residual-based multi-source fitting procedure to find the source components required by the data \citep{Alfaro2026_4HWC}.  Throughout this work, we interpret the fitted components as a likelihood-based decomposition of a crowded TeV field.

\subsection{Galactic Diffuse Emission Baseline}

Since this field lies on the Galactic plane, our model must account for Galactic diffuse emission. We use an energy-dependent diffuse template generated with High-Energy Radiative Messengers (HERMES), which simulates large-scale gamma--ray emission from cosmic rays interacting with interstellar gas and radiation fields \citep{Dundovic2021}. 
Following the 4HWC catalog analysis, the template includes the two processes expected to dominate at HAWC energies: IC scattering and $\pi^0$-decay emission \citep{Alfaro2026_4HWC}.
We include the HERMES diffuse template in the likelihood model with a fixed spatial and spectral shape and scale its overall normalization by the parameter $k_{\rm GDE}$
\citep{Dundovic2021, Alfaro2026_4HWC}. In the main analysis, we set $k_{\rm GDE}=1$, corresponding to the nominal HERMES prediction. This choice
provides a stable baseline model and limits how much the diffuse component can absorb flux from nearby source-like structures. To estimate the systematic effect of how we treat the diffuse background, we repeat the fits with $k_{\rm GDE}$ allowed to vary. We use the fits with $k_{\rm GDE}$ free only as systematic checks, not to define a separate source model. They show how the fitted source positions, extensions, and spectra change when the diffuse level can be adjusted. This comparison is especially important in the J1928 field, where extended source emission overlaps with structured Galactic-plane emission.

\subsection{Iterative Source Finding}

Using the reference analysis configuration, with the HERMES diffuse template included and fixed at $k_{\rm GDE}=1$, we build the source model through an iterative source-finding process. This follows the automated multi-source fitting approach from the 4HWC catalog and recent HAWC studies \citep{Alfaro2026_4HWC, Alfaro2024Boomerang}. 

At each iteration, we test the highest-significance residual as a new source candidate and refine it by testing its spatial shape and spectrum.
Each candidate is initially modeled as a point source with a power-law
(PL) spectrum of the form
\begin{equation}
\frac{dN}{dE} = K \left(\frac{E}{E_0}\right)^{-\alpha},
\label{eq:pl}
\end{equation}
where $K$ is the differential-flux normalization, $\alpha$ is the photon index, and $E_0$ is the pivot energy. As in the automated pipeline used in the 4HWC catalog \citep{Alfaro2026_4HWC}, we set the pivot energy to $E_0=2$~TeV for both the PL and log-parabola fits.

We compare nested models using the likelihood-ratio test statistic:
\begin{equation}
TS = 2 \left( \ln L_{\rm alt} - \ln L_{\rm null} \right),
\label{eq:ts}
\end{equation}
where $L_{\rm alt}$ and $L_{\rm null}$ are the likelihood values for the alternative and null hypotheses. We accept a new source into the model if it satisfies $TS>25$. We repeat this process until no additional candidate satisfies the detection threshold.

After establishing the source list, we test each significant source for symmetric Gaussian extension and spectral curvature. For the curvature test, we replace the PL model with a log-parabola (LP):
\begin{equation}
\frac{dN}{dE} = K \left(\frac{E}{E_0}\right)^{-\alpha-\beta \ln(E/E_0)},
\label{eq:lp}
\end{equation}
where $\alpha$ is the spectral slope at the pivot energy, and $\beta$ measures the curvature. 
We accept the extension or curvature when the model improves the fit by $\Delta TS>16$ relative to the simpler model. After each accepted change, we refit all free parameters to obtain the best-fit model of the region.

During the iterative source-finding process, we remove any candidate components that do not meet the detection threshold after the global refit, and we update the model without them. The process ends when no remaining candidates meet the detection threshold of $TS>25$. 
We call the resulting likelihood model the global source model. This model includes the fixed diffuse emission component with
$k_{\rm GDE}=1$ and all source components found by the iterative search, along with their fitted positions, shapes, and spectra.

After establishing the global source model, we determine a source-specific pivot energy for each spectral component.
We then repeat the final likelihood fit using these pivot energies.
In this step, we keep the pivot energies fixed. The pivot energy, $E_0$, is the reference energy where the normalization $K$ is defined.
Choosing $E_0$ close to the decorrelation energy helps reduce the covariance between the normalization and the spectral-shape parameters.
This choice also minimizes the relative uncertainty on the differential flux.
This procedure follows previous HAWC spectral analyses when reporting flux normalizations at energies within the source's sensitive energy range \citep{Albert2024GCPeVatron}.
The final pivot energies and spectral parameters are listed in Table~\ref{tab:global_model}.

\subsection{Energy Range and Spectral Energy Distribution}

After selecting the final spatial and spectral model, we determine the energy range over which HAWC data constrain each source well. For each source, we construct likelihood profiles to determine the lower and upper energy limits. We define $E_{\min}$ and $E_{\max}$ as the energies where the log-likelihood ($\ln L$) drops by $\Delta\ln L=0.5$ from its peak, which matches the $1\sigma$ profile-likelihood interval for one parameter. This method matches recent HAWC spectral studies \citep{Alfaro2024SS433}. We call this interval the HAWC-sensitive energy range for the source.

Within this energy range, the main spectral result is the best-fit model from the global likelihood analysis. To visualize the spectrum and compare it with other instruments, we also derive bin-by-bin flux estimates. In each energy bin, we adjust the source normalization while keeping the other spectral parameters fixed at the global best-fit values \citep{Abeysekara2017Crab, Abeysekara2019Crab}. We place each flux point at the median energy of the corresponding bin, based on the HAWC energy response and the best-fit source spectrum. Bins with $TS_{\rm bin}<4$ in the bin-by-bin normalization fit are reported as upper
limits rather than flux points. 

\section{Global Source Model of the Region}
\label{sec:global_model}

The global source model obtained from the iterative source-finding procedure consists of the fixed HERMES diffuse component, with $k_{\rm GDE}=1$, and six source components within the region of
interest. These components are the pulsar-centered source (HAWC~J1928+1750) near PSR~J1928+1746, three components in the G54.1+0.3 complex—one point-like component (HAWC~J1930+1852) near PSR~J1930+1852 and two nearby extended sources (HAWC~J1927+1849 and HAWC~J1929+1837)—as well as the southern point-like component  (HAWC~J1931+1655), and a broad eastern component (HAWC~J1922+1640). The concentration of cataloged pulsars and TeV emission in this region further motivates a multicomponent regional model.
Figure~\ref{fig:roi_context}(a) shows the HAWC significance map for the larger analysis region, with the source components from the final global model overlaid. Figure~\ref{fig:roi_context}(b)  shows the central source-only residual map around PSR~J1928+1746 after subtracting the other fitted source components and the Galactic diffuse component.
The pulsars shown in the figure serve as positional context and are not modeled as individual HAWC components. Table~\ref{tab:global_model} provides the best-fit positions, extensions, and spectral parameters for these components.

\begingroup

\renewcommand{\arraystretch}{1.4}
\setlength{\tabcolsep}{2.5pt}

\newcommand{\assoccell}[1]{%
  \parbox[t]{0.19\textwidth}{%
    \raggedright
    \setlength{\parindent}{0pt}%
    \setlength{\parskip}{0pt}%
    \linespread{0.92}\selectfont
    \strut#1\strut
  }%
}
\begin{deluxetable*}{lcccccccl}[t!]
\tabletypesize{\scriptsize}

\tablecaption{Best-fit parameters for the source components in the global source model of the J1928 region. The final column lists nearby catalog detections and possible associations.
\label{tab:global_model}
}
\tablewidth{0pt}
\tablehead{
\colhead{Source} &
\colhead{R.A.} &
\colhead{Dec.} &
\colhead{Extension ($\sigma$)} &
\colhead{$K$} &
\colhead{$\alpha$} &
\colhead{$\beta$}&
\colhead{$E_0$}&
\colhead{\shortstack{Nearby detections}}\\
\colhead{} &
\colhead{(deg)} &
\colhead{(deg)} &
\colhead{(deg)} &
\colhead{$(10^{-15}\,\mathrm{TeV}^{-1}\,
           \mathrm{cm}^{-2}\,\mathrm{s}^{-1})$} &
\colhead{} &
\colhead{} &
\colhead{(TeV)} &
\colhead{}
}

\startdata
HAWC~J1928+1750 & $292.13\pm0.02$ & $17.85\pm0.02$ & $0.19\pm0.02$ & $5.0\pm0.5$ & $2.09\pm0.11$ & $0.35\pm0.08$ & 12 & 
\assoccell{
3HWC~J1928+178;
1LHAASO~J1928+1746u
}
\\
HAWC~J1930+1852 & $292.59\pm0.02$ & $18.88\pm0.02$ & \nodata & $0.25_{-0.04}^{+0.05}$ & $2.54\pm0.09$ & \nodata & 19 & \assoccell{
HESS~J1930+188;
VER~J1930+188;
3HWC~J1930+188
} \\
HAWC~J1927+1849 & $292.00\pm0.03$ & $18.82\pm0.03$ & $0.20\pm0.03$ & $0.38^{+0.08}_{-0.06}$ & $2.38\pm0.14$ & $0.37\pm0.01$ & 30 & \assoccell{
1LHAASO~J1929+1846u
(KM2A)
}\\
HAWC~J1931+1655 & $292.81\pm0.02$ & $16.92\pm0.02$ & \nodata & $0.21^{+0.04}_{-0.03}$ & $2.45\pm0.10$ & \nodata & 20 & \assoccell{
1LHAASO~J1931+1653u
(KM2A)
}
\\
HAWC~J1922+1640 & $290.52\pm0.11$ & $16.07\pm0.11$ & $1.06\pm0.08$ & $380\pm40$ & $2.67\pm0.04$ & \nodata & 3 & \assoccell{
$\ast$3HWC~J1923+169;
1LHAASO~J1924+1609u
}
\\
HAWC~J1929+1837 & $292.43\pm0.07$ & $18.62\pm0.07$ & $0.92\pm0.07 $ & $283^{+30}_{-27}$ & $2.63\pm0.06$ & $0.19\pm0.04$ & 4 & \assoccell{
1LHAASO~J1929+1846u
(WCDA)
}
\\
\enddata

\tablenotetext{}{Coordinates are equatorial J2000 coordinates. The normalization $K$ is given at the source-specific pivot energy $E_0$ listed in the table. 
For power-law sources, $\alpha$ is the photon index. For log-parabola sources, $\alpha$ is the spectral slope at the pivot energy $E_0$, and $\beta$ gives
the curvature. The curvature parameter is not applicable for power-law sources.
The extension is the Gaussian width $\sigma$; point-like sources have no fitted extension.
Uncertainties are statistical only. 
The broad extended components in this table represent phenomenological likelihood components and should not always be interpreted as single, separate physical sources.
The nearby detections are taken from the 3HWC catalog \citep{Albert2020_3HWC}, the H.E.S.S. Galactic Plane Survey
\citep{Abdalla2018HGPS}, the VERITAS study of the G54.1+0.3 region
\citep{VERITAS_J1930}, and the first LHAASO catalog \citep{Cao2024LHAASO}.
These detections were selected within $0.5^\circ$ of the fitted HAWC centroid. If a catalog source met this criterion for multiple fitted components, it was assigned to the closest one. The source marked with an asterisk lies outside the nominal $0.5^\circ$ matching radius but remains within the fitted extension of the corresponding HAWC component. 3HWC J1923+169 \citep{Albert2020_3HWC} is approximately \(0.9^\circ\) from the centroid of HAWC J1922+1640 but lies within its fitted extension.}
\end{deluxetable*}

\endgroup

This model provides a more localized decomposition of the TeV emission than the earlier HAWC study of the J1928 region. The previous analysis used a shorter exposure ($\sim$ 1500 days), the $f_{\rm Hit}$ energy estimator, and no explicit Galactic diffuse emission template \citep{Albert2023J1928}. With the updated reconstruction, improved event selection, and the diffuse emission model used here, the broad central emission is redistributed
among HAWC~J1928+1750, the nearby TeV components, and Galactic diffuse emission.

This change is most significant near PSR~J1928+1746. In the previous HAWC analysis, the emission in this area was described by both an extended source, 3HWC~J1928+178, and a broader component, HAWC~J1928-EXT \citep{Albert2023J1928}. In the current fit, the pulsar-associated component
HAWC~J1928+1750 accounts for most of the peak emission near PSR~J1928+1746, while the remaining large-scale flux is modeled by nearby sources and the Galactic diffuse background.
A similar update occurs in the G54.1+0.3 region, where the new model separates HAWC J1930+1852 from the extended surrounding emission traced by HAWC J1927+1849 and HAWC J1929+1837. The point-like component HAWC~J1930+1852, discussed below, serves as a valuable cross-check for this decomposition.

\subsection{Cross-checks of the Global Source Model}

\label{sec:j1930_crosscheck}

\begin{figure}[t!]
\centering
\includegraphics[width=\columnwidth]{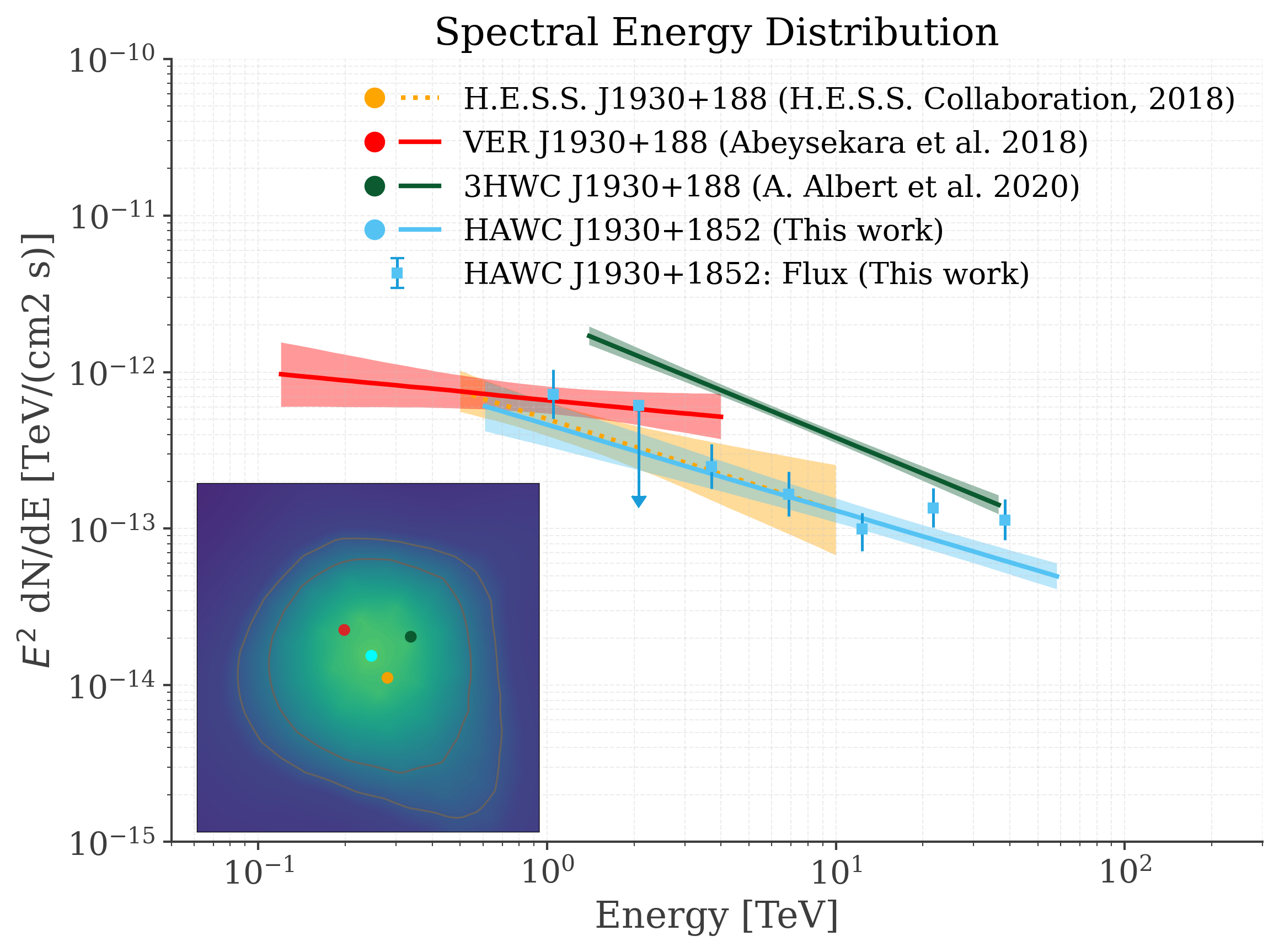}
\caption{Spectral comparison of HAWC~J1930+1852 with previous measurements of the
compact G54.1+0.3 source by H.E.S.S. \citep{Abdalla2018HGPS}, VERITAS
\citep{VERITAS_J1930}, and HAWC \citep{Albert2020_3HWC}. The spectrum from this
work agrees with the H.E.S.S. result over their common energy range and
extends to approximately 60~TeV. The inset shows a source-only residual significance map centered on HAWC J1930+1852, obtained after subtracting the other fitted source components and the Galactic diffuse-emission model while retaining HAWC J1930+1852. The cyan, green, orange, and red markers show the positions measured in this work, 3HWC, H.E.S.S., and VERITAS, respectively. The inset corresponds to the smaller white box in Figure \ref{fig:roi_context}(a).
}

\label{fig:j1930_sed}
\end{figure}

Before interpreting HAWC J1928+1750, we assess whether the six-component global model adequately describes this crowded region of interest. The six identified sources are listed in Table \ref{tab:global_model}. HAWC J1930+1852 provides the strongest external cross-check as it can be compared directly with independent IACT measurements. We then
place the remaining fitted components in the context of previous wide-field TeV detections and examine the post-fit residual-significance distribution.

HAWC J1930+1852 lies in the G54.1+0.3 region and matches the point TeV source detected near PSR
J1930+1852 and SNR G54.1+0.3 by H.E.S.S. and VERITAS \citep{Leahy2008, Abdalla2018HGPS,VERITAS_J1930}. The best-fit position from this work, $(\mathrm{R.A.},\mathrm{Decl.})=(292.59^\circ,18.88^\circ)$, agrees well with the reported localizations of HESS~J1930+188 and VER J1930+188. This is shown in the inset of Figure \ref{fig:j1930_sed}, which provides a
zoomed-in view of the left boxed region indicated in Figure \ref{fig:roi_context}(a).  As shown in Figure~\ref{fig:j1930_sed}, the HAWC spectrum agrees with the H.E.S.S. measurement over their common energy range and extends the point source spectrum to about 60 TeV. This agreement is notable because HAWC and H.E.S.S. use different observing techniques, have
different angular resolutions, and employ different methods to estimate the background. The consistency in both
position and spectrum supports the reliability of the global source model for this region.

The comparison with VERITAS and earlier HAWC catalog measurements should be interpreted with care because the source definitions, integration regions, and diffuse-background treatments differ. In the analysis presented here, the G54.1+0.3 emission is separated into the point-source component HAWC J1930+1852 and two nearby extended components, HAWC J1927+1849 and HAWC J1929+1837, with the HERMES Galactic diffuse-emission
template also included. Differences with the 3HWC spectrum are expected because 3HWC was a catalog-level analysis using 1523 days of data \citep{Albert2020_3HWC}, a previous reconstruction pass \citep{Smith2016HAWC, Abeysekara2017_2HWC, Abeysekara2018DAQ,
Abeysekara2019Crab}, and a search procedure that identified sources using a significance-valley source-separation criterion in point- and extended-source
optimized maps \citep{Albert2020_3HWC}. 
In contrast, the work presented here uses an updated reconstruction \citep{Albert2024HAWCPerformance} and a dedicated multicomponent model fit to characterize the total emission across the region.
Therefore, the 3HWC flux is not a one-to-one component comparison.

The remaining source components HAWC~J1927+1849, HAWC J1929+1837, HAWC J1922+1640, and HAWC~J1931+1655 are associated with previous detections by wide-field instruments: HAWC itself and LHAASO; see Table \ref{tab:global_model}. Differences between the sources presented here, and previous HAWC or LHAASO detections are either not
statistically significant, or arise from differences in specific source locations and extents, as well as in the GDE
modeling. Broad components like HAWC J1922+1640 and HAWC J1929+1837 are particularly sensitive to GDE
parameterizations, where varying background assumptions cause fitted centroids to shift, morphological extents to change, or spectral components to redistribute large-scale excess. These very extended sources likely represent a
composite mix of unresolved weak sources, structured diffuse emission, and source confusion. Notably, slight
variations in the global model evaluated in our systematic studies (see Section \ref{sec:systematics} below) do not significantly affect the measured properties of the TeV halo candidate HAWC J1928+1750, the primary subject of this paper.

\begin{figure}[t]
\centering
\includegraphics[width=\columnwidth]{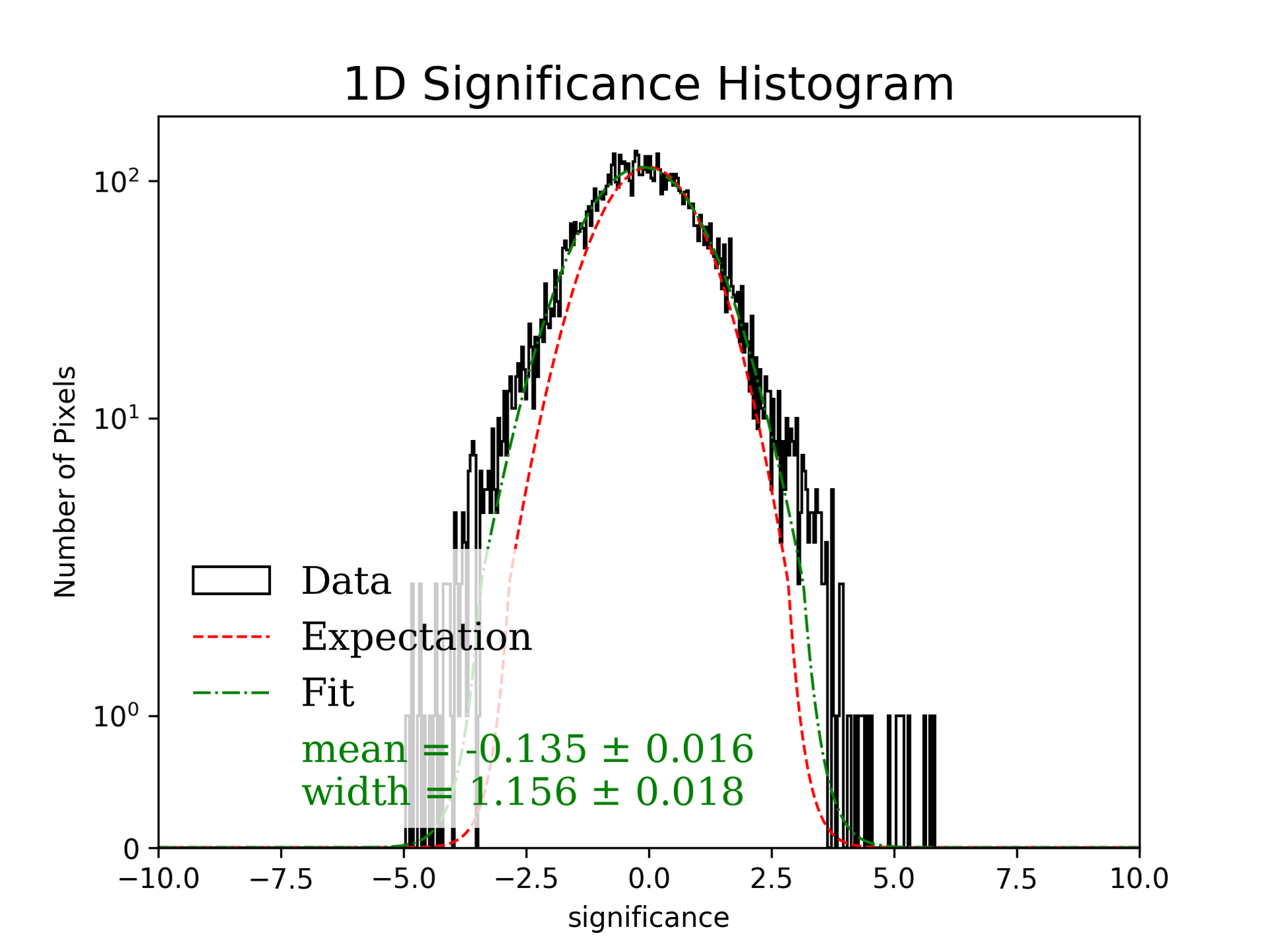}
\caption{
One-dimensional significance histogram of the post-fit residual map after subtracting the global source model, consisting of the six fitted source components and the HERMES Galactic diffuse component fixed at
$k_{\rm GDE}=1$. The histogram is computed within the analysis region centered on HAWC~J1928+1750 at $(\mathrm{R.A.},\mathrm{Decl.})=(292.13^\circ, 17.85^\circ)$.
}
\label{fig:residual_hist}
\end{figure}

As a final diagnostic of the global source model, we examine the residual-significance distribution after subtracting the six fitted source components and the HERMES Galactic diffuse component. Figure~\ref{fig:residual_hist} shows that the distribution is centered close to zero but broader than the unit-Gaussian expectation, with low-level tails on both sides. These deviations may reflect unresolved emission or residual background and emission-model mismatches in this crowded Galactic-plane field. No additional positive residual candidate reaches the source-finding threshold of $TS>25$. Therefore, we adopt the global source model as the reference baseline for the source-specific analysis of HAWC~J1928 + 1750.

\section{The TeV-Halo Candidate near PSR~J1928+1746}
\label{sec:halo}

This section examines HAWC~J1928+1750, the main TeV source near PSR~J1928+1746. The fitted HAWC centroid lies $0.04^\circ$ from the pulsar. This separation is consistent with the localization uncertainties and is small relative to the measured source extension. Figure~\ref{fig:roi_context}(b) shows
the central area around the pulsar, and Table~\ref{tab:source1} summarizes the best-fit
parameters of HAWC~J1928+1750, including both statistical and systematic uncertainties. Section~\ref{sec:systematics} describes the procedure used to estimate the systematic uncertainties.

Within the global source model, HAWC~J1928+1750 is represented by a symmetric Gaussian morphology and a log-parabola spectrum. We use the Gaussian model as an empirical description of the source shape, and test physically motivated interpretations later. The source is detected with high significance ($\sim 17 \sigma$) and is clearly extended, with a best-fit Gaussian width $\sigma \simeq 0.19^\circ$. Its fitted center is offset by only
$0.04^\circ$ from PSR~J1928+1746 and $0.05^\circ$ from the previously
reported 3HWC~J1928+178 position. The small centroid offset supports a spatial association with the pulsar environment \citep{Albert2020_3HWC, Albert2023J1928, Manchester2005}.

Compared to the previous HAWC analysis, the current model provides a more localized decomposition of the emission near PSR~J1928+1746. The earlier study used both an extended source and a broader component to model this region \citep{Albert2023J1928}. In this analysis, the emission near the pulsar is mainly described by the single, moderately extended component HAWC~J1928+1750, while the nearby TeV components and the Galactic diffuse background explain the remaining large-scale emission. 

\begin{figure}
\centering
\includegraphics[width=\columnwidth]{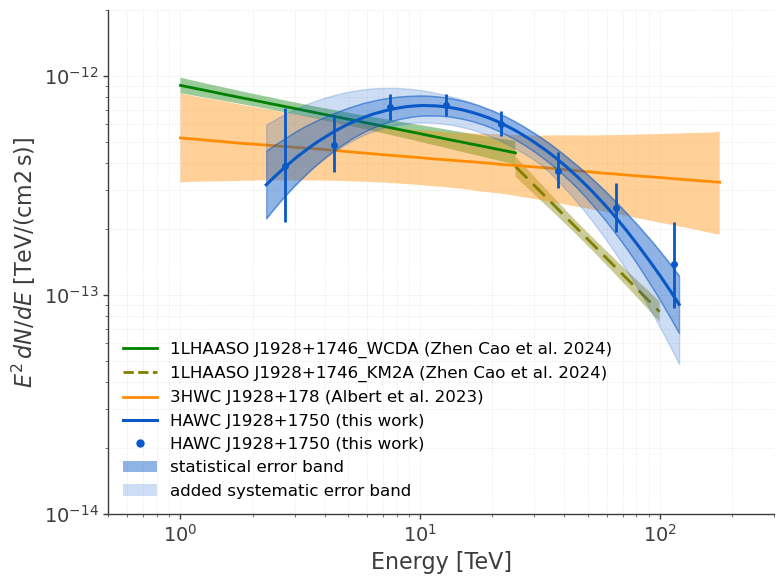}
\caption{Spectral energy distribution of HAWC~J1928+1750 from the reference analysis with $E_0=12$ TeV.
The blue points show the bin-by-bin HAWC flux estimates, and the solid blue curve shows the best-fit log-parabola model.
The darker and lighter blue bands indicate statistical uncertainty and the systematic envelope, respectively. 
The orange curve and band show the previous HAWC measurement of 3HWC~J1928+178 \citep{Albert2023J1928}, while the green and olive dashed curves show the WCDA and KM2A
measurements of 1LHAASO~J1928+1746u, respectively \citep{Cao2024LHAASO}. The \textit{Fermi}-LAT spectrum of J1929+1732E is not included here because the LAT source is broader than HAWC~J1928+1750.
}
\label{fig:j1928_sys_sed}
\end{figure}

The HAWC-sensitive energy range of HAWC~J1928+1750 extends from about 2~TeV up to energies exceeding 100~TeV, and its spectral energy distribution is presented in Figure~\ref{fig:j1928_sys_sed}. Compared with earlier HAWC measurements, the updated spectrum has smaller statistical uncertainties and a stronger fitted curvature, while overlapping the previous result within uncertainties over their common energy range \citep{Albert2023J1928}.

The updated HAWC spectrum is generally consistent with the published TeV measurements from LHAASO in this region. At lower TeV energies, the HAWC spectrum agrees with the 1LHAASO J1928+1746 (WCDA) measurement within uncertainties \citep{Cao2024LHAASO}. At higher energies, it is compatible with the 1LHAASO J1928+1746 (KM2A) detection, although direct comparison is complicated by structured emission in the region and by differences in angular resolution and spatial assumptions between the instruments. 

\begin{deluxetable}{lc}[t!]
\tabletypesize{\footnotesize}
\tablecaption{Best-fit parameters for HAWC~J1928+1750. Quoted uncertainties are given as statistical first and systematic second. For R.A. and Dec., the second entry gives the maximum absolute shift
of the fitted coordinate among the systematic tests; in both cases,
this shift is smaller than $0.02^\circ$.
\label{tab:source1}}
\tablewidth{0pt}
\tablehead{
\colhead{Parameter} &
\colhead{Value}
}
\startdata
R.A. (deg)        & $292.13 \pm 0.02\,$, $<0.02$ \\
Dec. (deg)       & $17.85 \pm 0.02\,$, $<0.02$ \\
Extension (deg)    & $0.19 \pm 0.02\,\,\pm 0.01\,$ \\
$K$ & $5.0 \pm 0.5\,\,{}^{+1.90}_{-0.61}\,$ \\
$\alpha$          & $2.09 \pm 0.11\,\,{}^{+0.25}_{-0.24}\,$ \\
$\beta$           & $0.35 \pm 0.08\,\,{}^{+0.08}_{-0.06}\,$ \\
\enddata
\tablenotetext{}{$K$ is quoted at 12~TeV in units of $10^{-15}\,\mathrm{cm^{-2}\,s^{-1}\,TeV^{-1}}$.}
\end{deluxetable}

The physical origin of HAWC~J1928+1750 remains uncertain, as previous studies have not identified a clear X-ray counterpart to the TeV emission near PSR~J1928+1746. Nearby molecular material provides possible target gas for hadronic gamma-ray production \citep{Albert2023J1928}. Thus, both a pulsar-powered leptonic scenario and a hadronic contribution from the surrounding environment remain possible.

To understand the source in the TeV-halo scenario, we analyze the emission near PSR~J1928+1746 using diffusion-halo templates based on physically motivated models. This approach helps us determine the diffusion coefficient value if electrons and positrons from the pulsar are causing the TeV emission. We create templates using Electron Diffusion and Gamma rays to the Earth (\texttt{EDGE}), following the methodology of \citet{LopezCoto2018EDGE} and similar studies conducted on Geminga and Monogem \citep{Abeysekara2017, Albert2024GemMon}.
In this model, particles injected by the pulsar propagate into the surrounding medium, lose energy through synchrotron and IC processes, and
produce TeV gamma rays via IC scattering on ambient radiation fields.

For the diffusion-template analysis, we adopt $B=8~\mu{\rm G}$ as the reference environmental assumption for the synchrotron-cooling calculation. This value is motivated by large-scale Galactic magnetic-field estimates for the inner Galaxy \citep{Beck2003MWB}, but it is not a direct measurement of the local magnetic
field around PSR~J1928+1746. Since such estimates are azimuthally averaged and
may not apply on the smaller scales relevant for individual TeV halos \citep{Seta2019Equipartition}, the inferred diffusion coefficient should be interpreted as conditional on this assumed magnetic field.

To estimate the diffusion coefficient, we start from the global source model described in Section~\ref{sec:global_model}. We construct a source-only residual map (Figure~\ref{fig:roi_context}(b)) from the reference HAWC NN data map by subtracting the fitted non-candidate source components and the Galactic diffuse component. The remaining residual emission is dominated by HAWC~J1928+1750. Next, we fit the diffusion-halo template to this residual map, keeping it centered on PSR~J1928+1746.

To double-check our results, we repeated the fit in the full multicomponent regional model. Here, the full multicomponent regional model refers to the complete global source model described in Section~\ref{sec:global_model}.
In this test, all fitted source components remain in the likelihood model, but the spatial morphology of HAWC~J1928+1750 is changed from the Gaussian used in the global source model to the diffusion-halo template centered on PSR~J1928+1746. 
This check assesses whether the inferred diffusion coefficient remains consistent when we model the surrounding sources and diffuse emission together rather than subtracting them before fitting the template.

\begin{deluxetable*}{lccc}
\tabletypesize{\scriptsize}
\tablecaption{Diffusion coefficient inferred for HAWC~J1928+1750
under the TeV-halo assumption using the $B=8~\mu{\rm G}$
diffusion-template set.
\label{tab:j1928_diffusion}}
\tablewidth{0pt}

\tablehead{
\colhead{Quantity} &
\colhead{Value} &
\colhead{Reference value} &
\colhead{Interpretation}
}
\startdata
$\log_{10}(D_0)$ & $25.81$ & -- & Fitted template parameter \\
$D_0$ (template normalization) & $6.46\times10^{25}$~cm$^2$~s$^{-1}$ & -- & Low-energy normalization \\
$D(100~{\rm TeV})$ & $2.1\times10^{27}$~cm$^2$~s$^{-1}$ & -- & Evaluated at $E_e=100~{\rm TeV}$ \\
$D(100~{\rm TeV})/D_{\rm ISM}$ & $\sim2.1\times10^{-3}$ & $D_{\rm ISM}\sim10^{30}$~cm$^2$~s$^{-1}$ & Lower than the ISM reference by a factor of $\sim500$ \\
$D(100~{\rm TeV})/D_{\rm Geminga}$ & $\sim0.66$ & $D_{\rm Geminga}\sim3.2\times10^{27}$~cm$^2$~s$^{-1}$ & Approximately $0.66$ of the Geminga central value \\
\enddata

\tablecomments{
The fitted $D_0$ is the normalization for the diffusion law used in the
template fit, $D(E_e)=D_0(1+E_e/3~{\rm GeV})^{1/3}$ \citep{LopezCoto2018EDGE}. The reported value of $D(100~{\rm TeV})$ is obtained by evaluating the diffusion law at $E_e=100~{\rm TeV}$. The templates use $B=8~\mu{\rm G}$ for the synchrotron-cooling calculation. The inferred diffusion coefficient should be interpreted as conditional on this assumed magnetic field. The Geminga reference is the central value from the individual-source fit of \citet{Abeysekara2017}. These comparisons provide reference scales, not tests of statistically significant differences.
}
\end{deluxetable*}

In the diffusion-template fit, the diffusion coefficient can vary within the template grid. The templates in this analysis use the following parameterization.
\begin{equation}
D(E_e)=D_0\left(1+\frac{E_e}{3~{\rm GeV}}\right)^\delta .
\end{equation}
$D_0$ refers to the low-energy normalization, and the fixed scale of 3~GeV follows the \texttt{EDGE} diffusion parameterization \citep{LopezCoto2018EDGE}. This scale marks the transition from energy-independent diffusion at low electron energies to $D(E_e)\propto E_e^\delta$ at high energies. The same convention is used in the template fit, and $\delta$ is fixed at 1/3, corresponding to a Kolmogorov-like energy dependence \citep{Kolmogorov1941, Strong2007}. The fitted parameter is $\log_{10}(D_0)$, with $D_0$ in ${\rm cm^2\,s^{-1}}$. For comparison with other TeV-halo studies, we evaluate the fitted
diffusion law at an electron energy of $100~{\rm TeV}$, and $D(100~{\rm TeV})$ is reported.

Table~\ref{tab:j1928_diffusion} presents the inferred diffusion coefficient based on the TeV-halo assumption and the set of diffusion templates $B=8~\mu{\rm G}$.
The converted value $D(100~{\rm TeV})\simeq2.1\times10^{27}~{\rm cm^2~s^{-1}}$ is much lower than the typical Galactic ISM value at similar energies,
$D_{\rm ISM}(100~{\rm TeV})\sim10^{30}~{\rm cm^2~s^{-1}}$
\citep{Strong2007}. It is also comparable in magnitude to the diffusion coefficient
inferred from the individual Geminga fit,
$D_{\rm Geminga}(100~{\rm TeV}) =
(3.2^{+1.4}_{-1.0})\times10^{27}~{\rm cm^2\,s^{-1}}$
\citep{Abeysekara2017}. Thus, the inferred diffusion is suppressed by a factor of $\sim500$ relative to the typical ISM value, consistent with the $\sim 100$--$1000$ suppression expected for TeV halos \citep{Evoli2018TeVHalos}. If the emission from HAWC~J1928+1750 comes from electrons and positrons escaping from PSR~J1928+1746, the data suggest that diffusion is suppressed around the pulsar.

\subsection{Radial Surface-Brightness Profile}
\label{sec:sb_profile}

To compare the empirical Gaussian morphology with the diffusion-halo template, we construct a radial surface-brightness profile for HAWC~J1928+1750. We center the profile on PSR~J1928+1746 because we are testing this source as a pulsar-powered TeV halo candidate. The best-fit TeV centroid is offset from
the pulsar by only about $0.04^\circ$, which is small compared to the measured Gaussian extension of the source. Pulsar proper motion could produce an offset between the current pulsar position and the center of relic TeV emission. However, no proper-motion vector is currently available for PSR~J1928+1746. We therefore center the profile on the pulsar as a physically motivated reference point.

\begin{figure}
\centering
\includegraphics[width=\columnwidth]{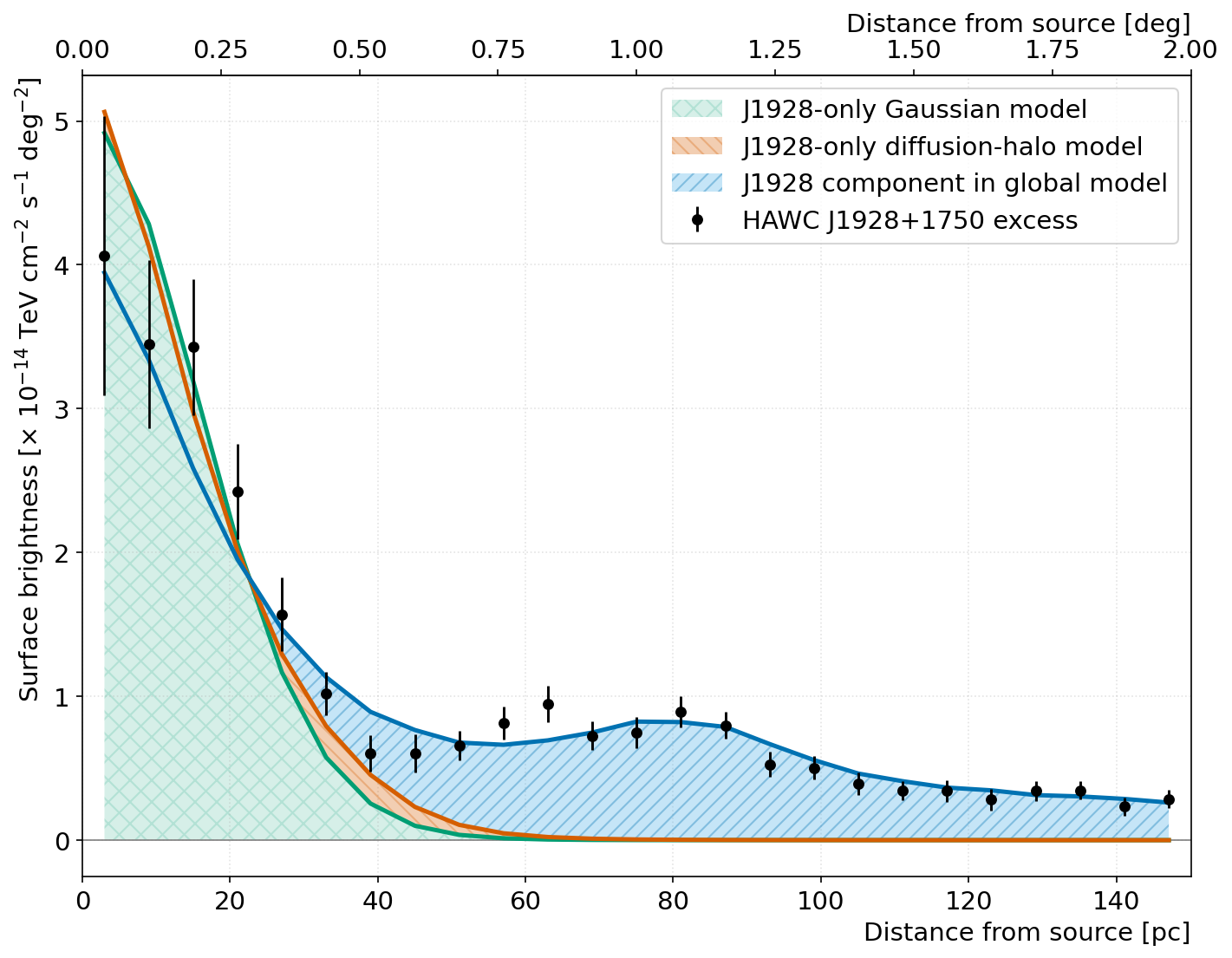}
\caption{
Surface-brightness profile around HAWC~J1928+1750, centered on
PSR~J1928+1746. Black points show the measured excess in concentric annuli.
The green and orange bands show the source-only expectations for HAWC~J1928+1750 from the Gaussian model and diffusion-halo template,
respectively. The blue band shows the HAWC~J1928+1750 component from the global model, not the sum of all model components. The colored bands represent model contributions, not
uncertainty bands. The lower axis gives the projected distance for the adopted distance
to PSR~J1928+1746, and the upper axis gives the corresponding angular separation. The inner emission is dominated by HAWC~J1928+1750, while the larger-radius emission is affected by nearby TeV components and Galactic-plane
emission.
}
\label{fig:sb_profile}
\end{figure}

Figure~\ref{fig:sb_profile} shows the measured surface-brightness profile together with the source-only expectations from the Gaussian
model and diffusion-halo template, as well as the
HAWC~J1928+1750 contribution obtained from the global source model. 
The colored bands represent model contributions, not uncertainty bands. Both the Gaussian and diffusion-halo models describe the compact inner emission associated with HAWC~J1928+1750. The similarity of the two profiles in the inner region is consistent with the small fitted diffusion coefficient, for which a slow-diffusion
halo can appear compact at HAWC's angular resolution.

At greater distances from the pulsar, the measured surface brightness exceeds the source-only expectations for HAWC~J1928+1750. This excess indicates that the outer profile is influenced by nearby TeV components and structured Galactic-plane emission. Therefore, HAWC~J1928+1750 dominates the bright inner emission near PSR~J1928+1746. The broader emission at larger radii should not be interpreted entirely as part of the halo candidate.

\subsection{Systematic Study}
\label{sec:systematics}

In addition to the statistical uncertainties from the nominal fit, we estimate systematic uncertainties for the best-fit parameters of HAWC~J1928+1750 using analysis-level cross-checks and changes in detector response, following standard HAWC procedures \citep{Abeysekara2019Crab, Albert2024HAWCPerformance}.
The nominal fit is defined as the fit of HAWC~J1928+1750 within the global source model, using the NN reconstruction and keeping the HERMES diffuse component fixed at $k_{\rm GDE}=1$.
The analysis-level checks look at the choice of energy reconstruction, either NN or GP, and how the Galactic diffuse model is handled, with
$k_{\rm GDE}$ set either as fixed or allowed to vary. The detector-response checks use alternative response files that probe potential detector mis-modeling, such as late-light effects, charge uncertainty, PMT threshold, and changes in PMT efficiency from run to run \citep{Albert2024HAWCPerformance}. For each parameter, we note the largest positive and negative differences from the nominal NN fit across these tests and combine the relevant contributions in quadrature to obtain a conservative systematic uncertainty.

Throughout these tests, the fitted centroid and extension of HAWC~J1928+1750 remain stable. Small changes in position and extension indicate that the detection and morphology of the source are reliable despite the changes in analysis methods and detector response considered here. However, systematic effects are more significant for the spectral parameters. Changes in energy reconstruction, diffuse background
treatment, and detector response can affect how the flux is divided among the pulsar-centered source, nearby components, and Galactic diffuse emission.

When we compare the NN and GP reconstructions,
we find that the Gaussian width changes from $0.19^\circ$ to $0.18^\circ$, a difference of about $4\%$. 
The GP fit raises the normalization by about $6\%$, steepens the spectral index by about $0.19$, and increases the curvature by about $0.03$. 
Allowing $k_{\rm GDE}$ to vary has a larger effect on the spectral decomposition. The normalization of HAWC~J1928+1750 increases by about $35\%$, and the spectral index and curvature also change compared to the
fixed-$k_{\rm GDE}$ fit. 
Changes in detector response cause further
spectral changes, with normalization varying by about $+13\%$ and $-12\%$, the spectral index changing by up to about $0.25$, and the curvature changing by up to about $+0.08$ and $-0.03$.

Overall, this systematic study shows that HAWC~J1928+1750 remains detected and that its fitted centroid and extension are stable. However, the detailed spectral
parameters are more sensitive to confusion with other sources and to how the diffuse background is modeled in this crowded region of the Galactic plane. Figure~\ref{fig:j1928_sys_sed} shows the reference HAWC spectrum for HAWC~J1928+1750, along with the statistical uncertainty band and the systematic envelope. For comparison, we also show previous TeV measurements of the J1928 region from HAWC and LHAASO.
Table~\ref{tab:source1} summarizes the best-fit values and uncertainties, with the first uncertainty being statistical and the second systematic.


\section{Scenarios for Gamma-Ray Production}
\label{sec:scenarios}

The TeV emission of HAWC~J1928+1750 is centered near PSR~J1928+1746, is spatially extended, and fits with a diffusion-halo model as described in Section~\ref{sec:halo}. These features point to a scenario in which the pulsar injects relativistic electrons and positrons that produce TeV gamma rays via IC scattering. The lack of a clear X-ray counterpart is consistent with an evolved PWN or TeV halo, where the synchrotron nebula may be faint, but IC emission is still detectable at TeV energies \citep{Albert2023J1928, Abeysekara2017, Albert2024GemMon}.

A hadronic contribution remains possible. The source is located in a complex region of the Galactic plane containing nearby molecular material, supernova-remnant candidates, and overlapping gamma-ray sources. Relativistic protons interacting with ambient gas could therefore contribute to the observed gamma-ray emission.
Previous multiwavelength studies of this region have considered both pulsar-powered scenarios and hadronic interpretations \citep{Albert2023J1928, Mori2020J1928}. In the present analysis, using a Galactic diffuse-emission template and a multicomponent TeV source model reduces the need for broad empirical components near PSR~J1928+1746, but they do not rule out hadronic emission from the surrounding environment.

To compare the potential emission channels, we perform phenomenological spectral modelling using \texttt{Naima} \citep{Zabalza2015}; these fittings are applied to the TeV SED after the HAWC likelihood analysis and are separate
from the empirical PL and log-parabola photon spectra used in the global source model. In the \texttt{Naima} fits, the parent electron or proton population is described by an exponentially cutoff power law:
\begin{equation} 
\frac{dN_i}{dE_i} = A_i \left(\frac{E_i}{E_{0,i}}\right)^{-\alpha_i} \exp\left(-\frac{E_i}{E_{{\rm cut},i}}\right).
\label{eq:naima_ecpl} 
\end{equation} 
Here, i=e for electrons/positrons and i=p for protons, $A_i$ is the normalization, $\alpha_i$ is the particle spectral index, and $E_{{\rm cut}, i}$ is the exponential cutoff energy.

In the leptonic model, gamma rays are produced through IC scattering by relativistic electrons and positrons. We include the cosmic microwave background, far-infrared, and near-infrared photon fields as target radiation. Since we do not include radio or X-ray synchrotron data in the likelihood, the magnetic-field assumption does not affect the gamma-ray spectral fit. In the hadronic model, gamma rays result from neutral-pion decay after proton-proton interactions. We apply the \citet{Kafexhiu2014} hadronic emission model as implemented in \texttt{Naima}.

\begin{figure}[t]
\centering
\includegraphics[width=\columnwidth]{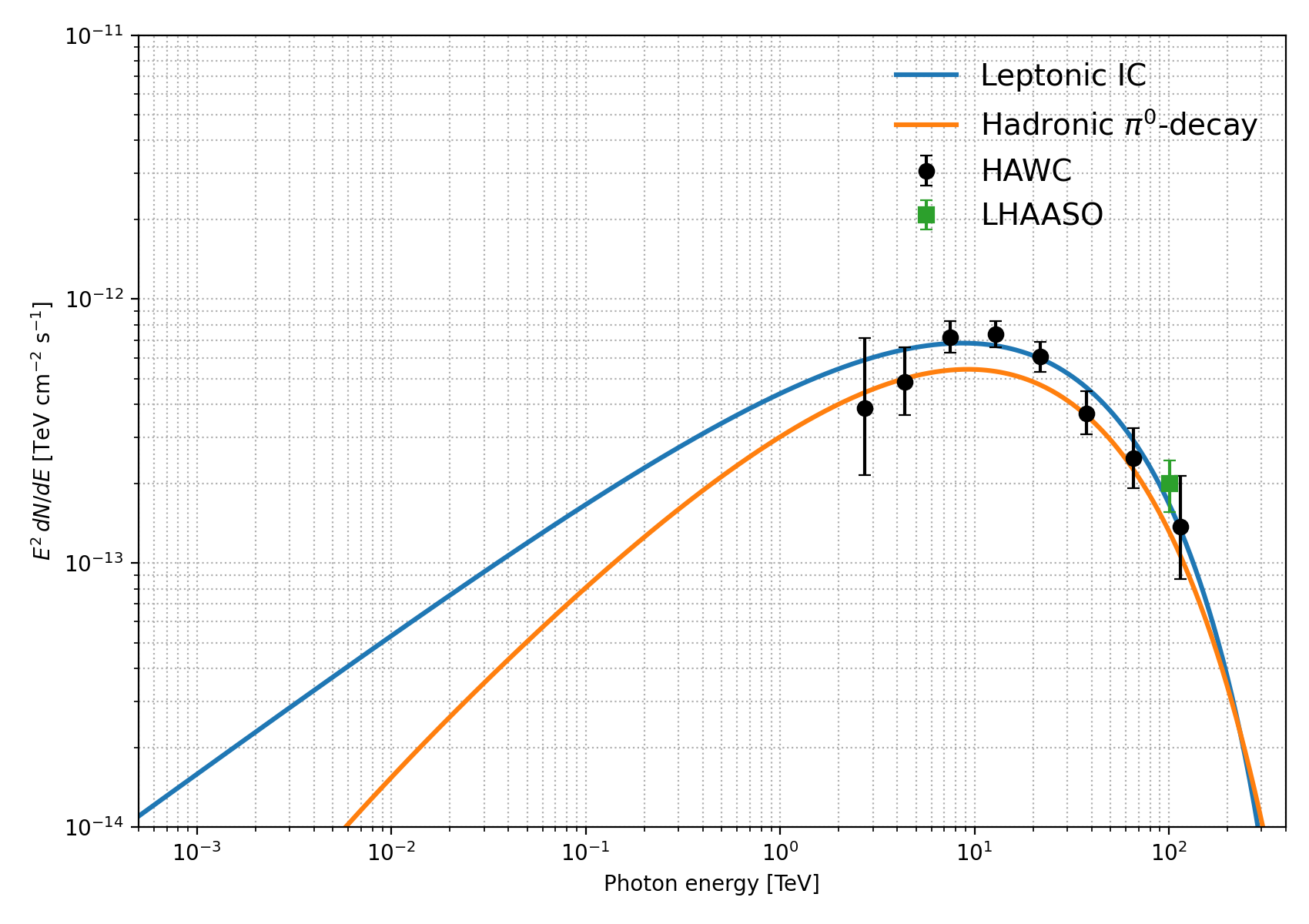}
\caption{
Phenomenological \texttt{Naima} package to fit the TeV spectrum of HAWC~J1928+1750. The models were applied only to the TeV data, based on the HAWC spectrum from this study and the LHAASO measurement. The LHAASO point is taken directly from the published LHAASO flux measurement for this source \citep{Cao2024LHAASO}. Both the one-zone leptonic IC model and the one-zone hadronic $\pi^0$-decay model fit the TeV SED equally well. This means the TeV spectrum alone cannot tell us whether the emission is leptonic or hadronic. We did not include the broader \textit{Fermi}-LAT component J1929+1732E \citep{Liu2026FermiJ1928} in this fit because it is more extended than HAWC~J1928+1750 and is not considered a direct GeV counterpart of the HAWC source.
}
\label{fig:naima_tev_only}
\end{figure}

Here, we fit only the TeV data using the HAWC spectrum from this work and the published LHAASO flux point for this source at higher energy \citep{Cao2024LHAASO}. Figure~\ref{fig:naima_tev_only} shows the one-zone IC and $\pi^0$-decay models, and Table~\ref{tab:naima_tev_only} lists the fitted parameters. For the leptonic model, the fit requires
$W_e(E_e>1~{\rm TeV})=3.31^{+2.94}_{-1.27}\times10^{46}~{\rm erg}$. For comparison, we estimate the characteristic spin-down energy ($E_{\rm sd}$) scale for PSR~J1928+1746 using its current spin-down power, $\dot{E}$, and characteristic age, $\tau_c$: \begin{equation}
E_{\rm sd} \gtrsim \dot{E}\tau_c \simeq 4.1\times10^{48}~{\rm erg}. 
\end{equation} 

This estimate does not include the full spin-down history of the pulsar, but it gives a useful conservative scale for comparison.
The estimated electron energy is less than about $1\%$ of this value, so the leptonic scenario is energetically reasonable for an evolved PWN or TeV halo \citep{Mori2020J1928, Liu2026FermiJ1928}.

To check our results, we estimate the average energy density of the TeV-emitting electrons using the method from \citet{Giacinti2020TeVHaloFraction}. For a
two-dimensional Gaussian morphology, the 68\% containment radius is $R_{68}\simeq1.51\sigma d$. Using $\sigma\simeq0.19^\circ$ and $d=4.3~{\rm kpc}$ gives $R_{68}\simeq22~{\rm pc}$. Combining this radius with
the electron energy inferred from the TeV-only leptonic fit, $W_e(E_e>1~{\rm TeV})\simeq3.3\times10^{46}~{\rm erg}$, gives
\[\epsilon_e \simeq \frac{W_e}{(4\pi/3)R_{68}^3}
\sim 2\times10^{-2}~{\rm eV~cm^{-3}} .
\]
This value is lower than the $\epsilon_e\sim0.1~{\rm eV~cm^{-3}}$ level used by \citet{Giacinti2020TeVHaloFraction} to identify low-energy-density, halo-like systems. So, along with the suppressed diffusion coefficient, the estimate of electron energy density supports a TeV-halo.
This estimate is only approximate because it depends on the assumed source radius, distance, and electron energy
range.

For the hadronic model, the TeV-only fit requires $W_p(E_p>1~{\rm TeV})=5.80^{+3.33}_{-1.08}\times10^{47}~{\rm erg}$. 
We use $n_{\rm H}=50~{\rm cm^{-3}}$ and $d=4~{\rm kpc}$, based on the molecular-cloud interpretation from \citet{Albert2023J1928}. That study used $^{13}{\rm CO}$ data from the Galactic Ring Survey to identify molecular
material overlapping the HAWC emission. The brightest velocity component, at $v_{\rm LSR}\sim22~{\rm km~s^{-1}}$, was divided into two clumps within the $>5\sigma$ HAWC emission region. We assume a cloud distance of $d=4~{\rm kpc}$, which is the most likely value from the \citet{Reid2016} Bayesian distance estimate; \citet{Albert2023J1928} calculated the clump
masses and volumes and found a typical gas density of $n_{\rm H}\sim50~{\rm cm^{-3}}$. The required proton energy scales roughly as $W_p\propto d^2/n_{\rm H}$, so if the average gas density were lower, the hadronic interpretation would require more energy.
This hadronic estimate represents the energy needed for a population of relativistic protons interacting with nearby gas. Molecular
material provides possible target gas, but it does not by itself identify the accelerator of those protons. Nearby SNR candidates including
G53.07+0.49 and G53.41+0.03, are present in the field \citep{Anderson2017THOR}, but their physical association with the compact
HAWC~J1928+1750 emission is not established. The hadronic scenario remains possible, but it requires an additional hadron accelerator or shock region, whereas the leptonic scenario is more directly motivated by the nearby energetic pulsar PSR~J1928+1746.

We compare the fit quality using the Bayesian Information Criterion,
${\rm BIC}=k\ln n-2\ln\hat{L}$, 
where $k$ is the number of free parameters, $n$ is the number of fitted spectral points, and $\hat{L}$ is the maximum likelihood \citep{Schwarz1978BIC}. Lower BIC values are preferred, but should only be compared between models fitted to the same data set.

The TeV-only IC and $\pi^0$-decay fits have similar BIC values with ${\rm BIC}=11.10$ for the IC model and ${\rm BIC}=10.54$ for the $\pi^0$-decay model. 
The difference, $\Delta{\rm BIC}\simeq0.6$, is too small to show a clear preference. As a result, the TeV data alone cannot distinguish between one-zone leptonic and hadronic scenarios.

\begin{deluxetable*}{llcccc}
\tabletypesize{\scriptsize}
\tablecaption{Summary of one-zone \texttt{Naima} fits to the TeV data for
HAWC~J1928+1750.
\label{tab:naima_tev_only}}
\tablewidth{0pt}
\tablehead{
\colhead{Model} &
\colhead{Data used} &
\colhead{Particle index} &
\colhead{$E_{\rm cut}$} &
\colhead{$W(>1~{\rm TeV})$} &
\colhead{BIC}
\\
\colhead{} &
\colhead{} &
\colhead{} &
\colhead{(TeV)} &
\colhead{(erg)} &
\colhead{}
}
\startdata
Leptonic IC &
HAWC+LHAASO &
$1.98^{+0.42}_{-0.45}$ &
$87^{+73}_{-28}$ &
$3.31^{+2.94}_{-1.27}\times10^{46}$ &
$11.10$
\\
Hadronic $\pi^0$ decay &
HAWC+LHAASO &
$1.46^{+0.34}_{-0.30}$ &
$238^{+165}_{-66}$ &
$5.80^{+3.33}_{-1.08}\times10^{47}$ &
$10.54$
\enddata
\tablenotetext{}{
For the leptonic model, $W(>1~{\rm TeV})$ denotes the total electron and
positron energy above 1~TeV. For the hadronic model, it denotes the total
proton energy above 1~TeV and is quoted for an assumed target density of
$n_{\rm H}=50~{\rm cm^{-3}}$ and distance $d=4~{\rm kpc}$, following the
molecular-cloud scenario considered by \citet{Albert2023J1928}. Hadronic
energies scale approximately as $W_p\propto d^2/n_{\rm H}$. The listed values
are posterior medians, with uncertainties corresponding to the 16th--84th
percentile range.
}
\end{deluxetable*}

A recent \textit{Fermi}-LAT analysis separated the GeV emission in this area into several components and identified an extended LAT component called J1929+1732E \citep{Liu2026FermiJ1928}. This name is specific to that study and is not an official 4FGL catalog name. The component overlaps with the official catalog source 4FGL J1929.0+1729 \citep{Ballet2023_4FGLDR4}, PSR J1928+1746, the previously reported HAWC emission near 3HWC J1928+178, and the LHAASO source 1LHAASO J1928+1746u. However, J1929+1732E is spatially broader than HAWC~J1928+1750, so we do not consider its GeV spectrum as a direct match to the compact HAWC source and do not include it in the quantitative Naima fits shown here. 
The \textit{Fermi}-LAT spectrum of J1929+1732E increases below a few GeV and then drops off, similar to the pion-decay feature seen in some older supernova remnants that interact with molecular material \citep{Liu2026FermiJ1928}. 
In contrast, the HAWC and LHAASO TeV spectrum is harder, extends to much higher energies, and is associated with a more compact TeV component. These differences suggest that the GeV and TeV emission may come from different particle groups or from different areas within the same complex region. For this reason, we use the \textit{Fermi}-LAT result only to provide context for the broader GeV to TeV environment, not as a direct spectral constraint on HAWC J1928+1750.

In summary, the TeV-only data do not provide sufficient evidence to differentiate between leptonic and hadronic models. 
The leptonic IC interpretation is energetically plausible for PSR~J1928+1746, while the hadronic interpretation remains viable if sufficient target gas and a plausible hadron accelerator exist. The broader \textit{Fermi}-LAT GeV component J1929+1732E shows that the GeV-to-TeV environment is complex and might include another hadronic component. Since this GeV component is more extended than HAWC~J1928+1750, we do not treat it as a direct spectral match for the compact TeV source. The compact TeV emission is therefore most plausibly interpreted as a pulsar-powered leptonic process, but we cannot rule out some hadronic contribution from the surrounding environment.

\section{Conclusions}
\label{conclusion}

We analyzed 2904 days of HAWC data from the Galactic plane region near PSR J1928+1746. For this analysis, we used updated event reconstruction and improved event selection methods. We also applied a multicomponent likelihood model that includes both TeV source components and Galactic diffuse emission.
Our reference model describes the region using six TeV components, which helps to separate the different sources in this complex area. 
Compared to the previous HAWC analysis, our updated model offers a more precise description of the emission near PSR J1928+1746.
The new model redistributes large-scale flux among nearby sources and the structured Galactic diffuse emission.
This results in a more localized and accurate view of the emission.

The main source of interest, HAWC J1928+1750, is detected with high significance near PSR J1928+1746, from about 2 TeV to above 100 TeV.
This source is moderately extended, with a Gaussian width of about 0.19 degrees, and has a curved TeV spectrum described by a log-parabola.
Systematic checks show that the source position and size remain stable when we change the energy reconstruction, the treatment of diffuse emission, or the detector response. However, the detailed spectral parameters are more sensitive to source confusion and to how we model the diffuse background.

If we interpret the source as a TeV halo, fitting diffusion-halo templates to HAWC J1928+1750 gives a diffusion coefficient of about $2.1\times10^{27}~{\rm cm^2~s^{-1}}$ for the set of B = $8 \mu$ G templates. This value is well below typical interstellar diffusion coefficients at comparable energies and is consistent with inhibited diffusion around pulsars. The radial surface-brightness profile supports this picture by showing that the compact inner emission is dominated by HAWC~J1928+1750, while the broader large-radius emission is affected by nearby TeV components and Galactic diffuse emission.

Models of gamma-ray production show that the TeV spectrum can be explained by either leptonic IC emission or hadronic $\pi^0$-decay emission.
The leptonic scenario is energetically reasonable, requiring only a small fraction of the characteristic spin-down energy scale of PSR~J1928+1746. The hadronic scenario is also possible if there is enough dense target gas and a hadronic accelerator. The broader \textit{Fermi}-LAT component J1929+1732E shows that the surrounding GeV to TeV environment is complex and may include additional hadronic emission, but it is not considered here as a direct spectral counterpart of the compact HAWC~J1928+1750 source. 

In summary, HAWC~J1928+1750 appears to be a strong candidate for a TeV halo or evolved PWN associated with PSR~J1928+1746. Although a hadronic contribution from the
surrounding environment cannot be excluded, the compact TeV emission is most naturally interpreted as a pulsar-powered leptonic component. The observed morphology, surface-brightness profile, diffusion-template analysis, and the plausible leptonic scenario together make HAWC~J1928+1750 a convincing TeV-halo or evolved-PWN candidate and provide a new constraint on inhibited particle transport near PSR~J1928+1746.


\section*{acknowledgments}

We acknowledge the support from: the US National Science Foundation (NSF); the US Department of Energy Office of High-Energy Physics; the Laboratory Directed Research and Development (LDRD) program of Los Alamos National Laboratory; Consejo Nacional de Ciencia y Tecnolog\'{i}a (CONACyT), M\'{e}xico, grants LNC-2023-117, 271051, 232656, 260378, 179588, 254964, 258865, 243290, 132197, A1-S-46288, A1-S-22784, CF-2023-I-645, CBF2023-2024-1630, c\'{a}tedras 873, 1563, 341, 323, Red HAWC, M\'{e}xico; DGAPA-UNAM grants IG101323, IG100726, IN111716-3, IN111419, IA102019, IN106521, IN114924, IN110521 , IN102223; VIEP-BUAP; PIFI 2012, 2013, PROFOCIE 2014, 2015; the University of Wisconsin Alumni Research Foundation; the Institute of Geophysics, Planetary Physics, and Signatures at Los Alamos National Laboratory; Polish Science Centre grant, 2024/53/B/ST9/02671; Coordinaci\'{o}n de la Investigaci\'{o}n Cient\'{i}fica de la Universidad Michoacana; Royal Society - Newton Advanced Fellowship 180385; Gobierno de España and European Union-NextGenerationEU, grant CNS2023- 144099; The Program Management Unit for Human Resources \& Institutional Development, Research and Innovation, NXPO (grant number B16F630069); Coordinaci\'{o}n General Acad\'{e}mica e Innovaci\'{o}n (CGAI-UdeG), PRODEP-SEP UDG-CA-499; Institute of Cosmic Ray Research (ICRR), University of Tokyo. H.M. acknowledges support under grant number CBF2023-2024-1630. H.F. acknowledges support by NASA under award number 80GSFC21M0002. C.R. acknowledges support from National Research Foundation of Korea (RS-2023-00280210, RS-2026-25470489). We also acknowledge the significant contributions over many years of Stefan Westerhoff, Gaurang Yodh and Arnulfo Zepeda Dom\'inguez, all deceased members of the HAWC collaboration. Thanks to Scott Delay, Luciano D\'{i}az and Eduardo Murrieta for technical support.

\bibliographystyle{aasjournalv7.1}
\bibliography{references}

\end{document}